\documentclass[sigconf, nonacm]{acmart}
\usepackage{algorithm}
\usepackage{algorithmic}
\usepackage{graphicx}
\usepackage{float} 
\usepackage{subfigure}
\usepackage{array}
\usepackage{amsfonts}
\usepackage{enumitem}
\usepackage{color}
\usepackage{url}
\usepackage{extarrows}
\usepackage{multirow}
\usepackage{threeparttable}
\usepackage{bbm}
\usepackage[utf8]{inputenc}

\usepackage{cleveref}
\crefname{section}{§}{§§}
\Crefname{section}{§}{§§}
\newtheorem{problem}{Problem}

\newcommand{\revision}[1]{\textcolor{blue}{#1}}
\newcommand{\techreport}[0]{}
\newcommand{\paper}[1]{}
\newtheorem{example}{Example}

\newcommand\vldbdoi{XX.XX/XXX.XX}
\newcommand\vldbpages{XXX-XXX}
\newcommand\vldbvolume{14}
\newcommand\vldbissue{1}
\newcommand\vldbyear{2020}
\newcommand\vldbauthors{\authors}
\newcommand\vldbtitle{\shorttitle} 
\newcommand\vldbavailabilityurl{http://vldb.org/pvldb/format_vol14.html}
\newcommand\vldbpagestyle{plain} 

\AtBeginDocument{%
  \providecommand\BibTeX{{%
    \normalfont B\kern-0.5em{\scshape i\kern-0.25em b}\kern-0.8em\TeX}}}

\begin{document}
\paper{
{\Huge\textbf{Response to Reviewer Comments}\\}

Dear Reviewers:

  We are very grateful to all reviewers for taking the time to provide comments on our paper. We have revised our paper based on these valuable comments. The responses to the revision requests and the relevant revision actions taken are as follows.
  
  All revised text is in blue and specific changes are summarized below.\\
       
\vspace{-0.2cm}
{\LARGE\textbf{Summary of Major Changes}}

Most of the changes are concentrated in the first two sections, the experimental part, abstract and conclusion:
\begin{enumerate}[itemsep=5pt]
    \vspace{-0.1cm}
    \item We use the performance gap of RMIs on synthetic and real-world datasets as motivation to rewrite the first two sections of this paper. Specifically, we argue that the key reason for the performance degradation in practical scenarios is that existing RMIs ignore the importance of data partitioning. Therefore, we motivate this paper from this aspect. To deal with this issue, we explicitly apply the data partitioning view to the index construction through a new cost model. To enhance the flexibility of data partitioning during construction, we formalize the index construction problem as an optimization problem and solve it through a hybrid algorithm. Moreover, we also equip CARMI with a cache-friendly design to further improve the performance on real-world datasets. The organization of the paper has also been changed as a result. In addition, we have also added an overview section in Section 2.2 to help understand the basic structure of CARMI.
    \item In the experimental part, we have added another real-world dataset to better reflect the performance of indexes, and describe the experimental evaluation in detail in Section~\ref{sec:setup}. Furthermore, we have added a detailed analysis of the results on the SOSD platform in Section~\ref{sec:sosd} and the choice of the two main parameters of CARMI in Sections~\ref{sec:tradeoff} and~\ref{sec:cost}, respectively.
    \item To better illustrate the performance gains of different ideas, we have also added the performance breakdown in Section~\ref{sec:breakdown}. Specifically, we implement four variants of CARMI to simulate indexes equipped with different ideas. The results show that most of the performance gains come from the data partitioning view, and the cache-aware design and DP algorithm can also bring certain benefits to CARMI, as discussed in Section~\ref{sec:breakdown}.
    \item To examine the robustness of CARMI, we have added experiments with 10 different workload and data distribution shifts and discussed the results in detail in Section~\ref{sec:robust}. We divided them into three groups: access pattern shifts in read-only workloads, read-write workload shifts and data distribution shifts. Experimental results show that CARMI can still execute queries quickly and is robust in these cases.\\
\end{enumerate}

{\LARGE\textbf{Details of Changes\\}}

\vspace{-0.2cm}
{\Large\textbf{Reviewer \#1}}
\begin{enumerate}[itemsep=5pt]
    \item[\textbf{(D1)}] While it is understandable that not all the details of the technical report can make it into the paper, the experimental evaluation should provide enough detail for one to understand the core experiments without having to consult the technical report.
    
    \textbf{Response}: Thanks for your suggestion, we have moved some figures and discussions back into the paper and reduced references to technical reports. Some examples are as follows:

    \begin{enumerate}[itemsep=5pt]
        \item[(a)] Please brieﬂy describe the time metric reported in Figure 10 and how it is measured.

        \textbf{Response}: For each workload, we evaluate the total time of 100,000 queries, and report the average time used by each query. Keys to look up are selected randomly from the existing keys in the index and we use two access patterns: a Zipfian distribution and a uniform distribution. We have added a paragraph in Section~\ref{sec:workload} to describe the experimental evaluation in detail.
    
        \item[(b)] Section 6.1.2 says there are 5 different query workloads, but I only count 4. I do not see any results for the write-partial workload.

        \textbf{Response}: Thanks for your correction. Due to page limitations, we only show 4 workloads in the paper, and the results of the write-partial workload mentioned in the original text are in the technical report. Inconsistencies and typos in Section~\ref{sec:workload} have been fixed.
    
        \item[(c)] A description of the CARMI parameters is one of the most important discussions missing in the paper. You already provide some of this information in the technical report. Please include something similar in the paper and also explain how you chose the default settings and the amount of tuning they generally require. For the ``lambda'' and ``kAlgorithmThreshold'' parameters, I think it makes sense to discuss these details in Sections 6.4 and 6.5 as part of the analysis. Lastly, you should include any parameters you set for the baselines.

        \textbf{Response}: We have added a brief description of the parameters of CARMI and baselines in Section~\ref{sec:paramsSetting}. We tune the node size of the B+ Tree to 512 bytes, which is optimal for in-memory queries~\cite{zhang2016reducing}. ALEX does not require any tuning, as described in their paper~\cite{ding2019alex}. For CARMI, we only tune one parameter $\lambda$, and all other parameters use default values unless otherwise stated, as described in Section~\ref{sec:paramsSetting}. In Section~\ref{sec:tradeoff}, we discuss the effect of $\lambda$ in detail and demonstrate through experiments that this parameter needs to be carefully tuned based on the data distribution to obtain a more desired tradeoff between time and space. The $kAlgorithmThreshold$ parameter (which has been renamed to $kDPThreshold$) is discussed in detail in Section~\ref{sec:cost}. In general, this parameter does not need to be tuned, and the default value can be used directly, except in some scenarios where build time is very important.
    
        \item[(d)] Is the ``OSMC dataset" in Section 6.2 the same as the ``osmc64" dataset in Section 6.3? If they are, then in Section 6.2.1 you update the reference ``(the CDF can be found in our technical report[45])" to Figure 11.

        \textbf{Response}: Thanks for your correction. The reference in Section~\ref{sec:readonlyResult} has been updated to this figure.
    
        \item[(e)] There are two issues with Section 6.3: (1) it does not provide any details about the baselines and datasets considered in the experiments, and (2) there is almost no analysis of the results presented in Tables 3 and 4.

        \textbf{Response}: We use all real-world datasets (each dataset consists of 200 million unsigned 32-bit/64-bit integers) and baselines in SOSD, and detailed descriptions have been added in Section~\ref{sec:sosd}. We have also provided a more detailed analysis of the results in this section. Among them, CARMI can beat all the baselines on these datasets and the used space is comparable to that of traditional indexes with good performance.
    
        \item[(f)] Section 6.5 says Figure 13 shows index construction finishes within 1-4 minutes. What I see in Figure 13 is that the index construction time ranges from approximately 0.2 to 1.6 minutes. Is ``1-4 minutes" a typo? If not, can you clarify?
         
        \textbf{Response}: The ``4 minutes'' in the text also includes the construction time on real-world datasets.  The construction time of CARMI can be adjusted by the $kDPThreshold$ parameter. When the value of $kDPThreshold$ is 1024, it takes three minutes on the Facebook dataset. In most cases, the build time is around 1 minute or less. We have redrawn this figure to include real-world datasets and updated the text in Section~\ref{sec:cost} accordingly.
    \end{enumerate}
    \item[\textbf{(D2)}] When constructing an index, how do you choose the number of child nodes to consider for non-leaf nodes?

    \textbf{Response}: Thank you for pointing out the details that were lacking in our original description. The way to select the number of child nodes (in a list of powers of 2) has been added at the beginning of Section~\ref{sec:constructionAlg}.

    \item[\textbf{(D3)}] Section 6.2.1 mentions a ﬂatter index structure is one reason CARMI is faster than B+ Tree (due to fanout) and ALEX (at least for datasets with non-linear local CDFs). Can you share the average tree depth of each index in this discussion?

    \textbf{Response}: The average tree depth of CARMI is 2.1, while that of ALEX and B+ Tree are 2.5 and 7.3, respectively. We have added the average tree depth in Section~\ref{sec:readonlyResult}.

    \item[(D4)] Is it possible to update Figure 13 to demonstrate the tradeoff between build time and efficiency by also including the query latency?

    \textbf{Response}: We have updated this figure to include the query latency in Section~\ref{sec:cost}.

    \item[\textbf{(D5)}] Figures 12 and 13 only provide results for the synthetic datasets. Why are none of the real-world datasets were included?

    \textbf{Response}: These two figures have been redrawn to include real-world datasets in Section~\ref{sec:tradeoff} and~\ref{sec:cost}.

    \item[(D6)] There are opportunities to condense and/or remove some of the text and ﬁgures from Sections 1-5 so that the experimental evaluation and related work can be expanded.

    \textbf{Response}: Thank you very much for your suggestions and all of them have been adopted in this revised version. We have condensed the previous sections as you suggested and expanded the sections of experimental evaluation in Section~\ref{sec:experiment}, and related work in Section~\ref{sec:related}. In addition, we have rewritten the Introduction to make the logic of this paper more coherent and clear.

    \item[\textbf{(D7)}] The paper has a few typos and minor presentation issues.
    
    \textbf{Response}: Thanks for your correction. These typos have been fixed.\\

\end{enumerate}

\vspace{-0.2cm}
{\Large\textbf{Reviewer \#2}}
\begin{enumerate}[itemsep=5pt]
    \item[(D1)] The performance section is currently quite insuﬃcient; I see that you ran some numbers, but there is little intuition why performance is better. There is a bit of info in Section 6.3, but that is very vague. Please dig more deeply to understand where the performance improvements are coming from.
    
    \textbf{Response}: Thanks for your suggestion, we have revised Section~\ref{sec:experiment} to better analyze the experimental results. (1) We have added a performance breakdown in Section~\ref{sec:breakdown} to illustrate the contribution of each idea through four variants of CARMI equipped with different ideas. The results show that the data partitioning view contributes the most to CARMI. Our new cost model can well characterize the performance of different nodes and provide more flexible data partitioning. In addition, the cache-aware design and DP algorithm also bring some benefits to CARMI. (2) In Section~\ref{sec:sosd}, we add a description of the experimental evaluation and an analysis of the results, and make the results more obvious by boldening the data that performs well in Table~\ref{tab:sosdPaper}. (3) With respect to performance gains, we have rewritten the performance representation more clearly in the Introduction and also added another real-world dataset to better show the outstanding performance of CARMI. On the SOSD benchmark, CARMI can still achieve an average speedup of 1.2$\times$ compared to RMI, which has been carefully tuned for each dataset in advance.
    
    \item[(D2)] In Theorem 4.1, $C(M_i)$ is not deﬁned?
    
    \textbf{Response}: We have renamed the ambiguous symbols. $C(M_i)$ is renamed to $Capacity(M_i)$, representing the leaf node's capacity, as shown in Theorem~\ref{theo:entropy}.
    
    \item[(D3)] Would be great to see a table with the deﬁnition of all the terms.
    
    \textbf{Response}: We have renamed all variables in Section~\ref{sec:framework} and~\ref{sec:constructionAlg} to more intuitive names such as $Capacity(M)$. If the current names and definitions are still not clear enough, we will add a table and adjust the page space as you suggested in the next revision.\\
\end{enumerate}

\vspace{-0.2cm}
{\Large\textbf{Reviewer \#3}}
\begin{enumerate}[itemsep=5pt]
    \item[\textbf{(D1)}] The paper is hard to follow since it reads like a loosely coupled collection of ideas to improve the RMI framework which is not strongly connected to the motivation of the paper (cache-awareness, cost-based construction).
     
    \textbf{Response}: Thanks for your valuable suggestions. We have rewritten the paper and motivated it by the performance degradation of RMIs on real-world datasets and propose new ideas to solve it accordingly. We argue that existing RMIs are not flexible enough in data partitioning during index construction, resulting in performance degradation in real scenarios. Therefore, we rewrite the paper from this aspect, especially the first two sections.

    \item[\textbf{(D2)}] Related to D1, it would be natural to ask how much each individual contribution contributes to the overall performance gains. This would make it much easier to assess the overall contribution.

    \textbf{Response}: In addition to rewriting the paper, we add a detailed analysis of CARMI in Section~\ref{sec:breakdown} to demonstrate the contribution of each idea. Specifically, we implement four variants of CARMI to simulate indexes equipped with different ideas and perform read-only workloads on 5 datasets. The results show that most of the performance gains of the CARMI framework come from the data partitioning view. The new cost model based on entropy can well characterize the performance of different nodes, so that the index constructed only by the greedy algorithm can have a speedup of 1.84$\times$ compared to the static two-layer RMI. Moreover, the cache-aware design and DP algorithm can also bring certain benefits to CARMI, as discussed in Section~\ref{sec:breakdown}.

    \item[(D3)] Since cache-awareness is a major motivation, cache-aware variants of B-tree (CSB+ trees, etc.) should be evaluated as baselines.

    \textbf{Response}: Thanks for your suggestion. The STX B+ Tree we used is a modern version and is a state-of-the-art implementation of B+ Tree that already implements many cache-aware ideas in CSB+ Tree. For example, the node size is aligned with the cacheline, and keys and values are stored separately to better utilize the cache. In addition, the current implementation of CSB+ Tree is designed for 32-bit systems and only supports the $int$ type, which requires completely rewriting the code if we need to use it in our experiments. If CSB+ Tree is absolutely necessary, we will reimplement it and use it as a baseline in the next revision.

    \item[\textbf{(D4)}] The construction times of the index should be compared with the baselines. This is interesting since depending on the workload it might not be worth to spend this additional time on building the index. Again, minutes for merely one GB of data seem signiﬁcant.

    \textbf{Response}: We have added a comparison of the construction time of CARMI and that of the baselines in Section~\ref{sec:cost} and redrawn the figure to add latency. B+ Tree and ALEX take 10s and 20s respectively to construct indexes. The construction of CARMI can be finished within 0.3-3.1 minutes. Note that CARMI can adjust the construction time by tuning the $kDPThreshold$ parameter. If the construction time is very important, we can reduce this parameter to construct indexes faster and the performance will not drop too much. When this parameter is 96, CARMI takes only 19-40 seconds to construct indexes for these datasets.

    \item[(D5)] How well would an index perform that is only constructed greedily?

    \textbf{Response}: The latency of indexes only constructed greedily has been added in Section~\ref{sec:breakdown} and~\ref{sec:cost}. The results show that the performance difference between the index built with only the greedy algorithm and the index built with the hybrid construction algorithm is about 10 ns, while the construction time on the Facebook dataset can be reduced from 3.1 minutes to 33 seconds.

    \item[(D6)] In addition in your experiments, how did you set the kAlgorithmThreshold? Depending on this, construction times would be even more expensive and likely much higher than for the baselines.

    \textbf{Response}: The value of this parameter is the default value (512) in our experiments except in Section~\ref{sec:cost}. The construction time of most indexes is less than 1 minute and the specific results can be found in Figure~\ref{fig:time}.

    \item[\textbf{(D7)}] The index construction relies on exact knowledge of workload and data distribution. This is of course reasonable to assume but I wonder how robust the generated indexes are w.r.t. to (i) workload and (ii) data drifts.
    
    \textbf{Response}: We test the robustness of CARMI to workload and data distribution shifts and the experimental results have been added in Section~\ref{sec:robust}. Specifically, we test the robustness of CARMI in terms of access pattern shifts, read-write workloads shifts, and data distribution shifts. The results show that CARMI is robust to data distribution and workload shifts.
    
    (1) We construct indexes from evenly distributed historical queries on OSMC/Face datasets, and query them with 8 different access patterns: seven Zipfian distributions and a uniform distribution. The results show that CARMI still has good performance when new sets of keys are queried more often. (2) We construct indexes with historical queries under read-heavy and write-heavy workloads, respectively, and test them with write-heavy workloads. Although the index built on read-heavy workloads requires 1.1$\times$ the average time compared to the index built on write-heavy workloads, as shown in Figure~\ref{fig:drift}(a), CARMI still executes queries quickly. (3) We use the uniform dataset to build the index, and then perform a write-heavy workload using data points from OSMC dataset. Since inserted data points come from a different dataset, we can simulate the situation where the data distribution gradually changes from a uniform distribution to a non-linear distribution. Figure~\ref{fig:drift}(b) shows that although CARMI requires slightly more average time than the index built on OSMC dataset, CARMI can still maintain good performance in this situation.
\end{enumerate}

\thispagestyle{empty}
}

\setcounter{page}{1}
\title{CARMI: A Cache-Aware Learned Index with a Cost-based Construction Algorithm}

\author{Jiaoyi Zhang}
\affiliation{%
  \institution{Tsinghua University}
  \city{Beijing}
  \state{China}
}
\email{jy-zhang20@mails.tsinghua.edu.cn}

\author{Yihan Gao}
\affiliation{%
  \institution{Tsinghua University}
  \city{Beijing}
  \state{China}
}
\email{gaoyihan@mail.tsinghua.edu.cn}

\begin{abstract}
Learned indexes, which use machine learning models to replace traditional index structures, have shown promising results in recent studies. However, existing learned indexes exhibit a performance gap between synthetic and real-world datasets, making them far from practical indexes.

In this paper, we identify that ignoring the importance of data partitioning during model training is the main reason for this problem. Thus, we explicitly apply data partitioning to index construction and propose a new efficient and updatable cache-aware RMI framework, called CARMI. Specifically, we introduce entropy as a metric to quantify and characterize the effectiveness of data partitioning of tree nodes in learned indexes and propose a novel cost model, laying a new theoretical foundation for future research. Then, based on our novel cost model, CARMI can automatically determine tree structures and model types under various datasets and workloads by a hybrid construction algorithm without any manual tuning. Furthermore, since memory accesses limit the performance of RMIs, a new cache-aware design is also applied in CARMI, which makes full use of the characteristics of the CPU cache to effectively reduce the number of memory accesses. Our experimental study shows that CARMI performs better than baselines, achieving an average of $2.2\times$/$1.9\times$ speedup compared to B+ Tree/ALEX, while using only about $0.77\times$ memory space of B+ Tree. On the SOSD platform, CARMI outperforms all baselines, with an average speedup of $1.2\times$ over the nearest competitor RMI, which has been carefully tuned for each dataset in advance.

\end{abstract}

\thispagestyle{plain}


\maketitle
\pagestyle{\vldbpagestyle}
\begingroup\small\noindent\raggedright\textbf{PVLDB Reference Format:}\\
\vldbauthors. \vldbtitle. PVLDB, \vldbvolume(\vldbissue): \vldbpages, \vldbyear.\\
\href{https://doi.org/\vldbdoi}{doi:\vldbdoi}
\endgroup
\begingroup
\renewcommand\thefootnote{}\footnote{\noindent
This work is licensed under the Creative Commons BY-NC-ND 4.0 International License. Visit \url{https://creativecommons.org/licenses/by-nc-nd/4.0/} to view a copy of this license. For any use beyond those covered by this license, obtain permission by emailing \href{mailto:info@vldb.org}{info@vldb.org}. Copyright is held by the owner/author(s). Publication rights licensed to the VLDB Endowment. \\
\raggedright Proceedings of the VLDB Endowment, Vol. \vldbvolume, No. \vldbissue\ %
ISSN 2150-8097. \\
\href{https://doi.org/\vldbdoi}{doi:\vldbdoi} \\
}\addtocounter{footnote}{-1}\endgroup

\ifdefempty{\vldbavailabilityurl}{}{
\vspace{.3cm}
\begingroup\small\noindent\raggedright\textbf{PVLDB Artifact Availability:}\\
The source code, data, and/or other artifacts have been made available at \url{\vldbavailabilityurl}.
\endgroup
}

\setcounter{page}{1}
\section{Introduction}

As an indispensable access method of database systems, indexes provide fast data accesses by avoiding expensive table scans. Traditional index structures are general-purpose, in the sense that they organize data according to fixed rules without taking advantage of the characteristics of underlying data distribution. Recently, Kraska et al.~\cite{kraska2018case} pioneered a line of research where indexes are constructed using machine learning models. Specifically, they proposed a structure called the Recursive Model Index (RMI). In RMI, index nodes themselves are ML models, and they are connected hierarchically. To perform a lookup, we traverse the tree-like structure using the ML model in each node to determine the child branch to continue. Upon arriving at a position in the data array (i.e., leaf nodes), we perform the ``last-mile'' search within a range to correct the model prediction error. \cite{kraska2018case} demonstrates that the RMI and learned indexes in general exhibit smaller memory consumption and superior search performance compared to B+ Trees.

Despite the promising results shown in the latest research~\cite{kraska2018case,ding2019alex,radixspline,ferragina2020pgm,stoian2021plex,galakatos2018tree,marcus2020cdfshop}, few real database systems choose to adopt learned indexes. An outstanding reason, as indicated in the SOSD benchmark paper~\cite{marcus2020benchmarking}, is that the performance of learned indexes (represented by the RMIs) drops significantly when moving from synthetic datasets to real-world datasets. The average latency of an index lookup is 2.92$\times$ larger on real-world datasets than that on the same-sized synthetic ones, as shown in~\cite{marcus2020benchmarking}.

We argue that the main reason behind such a performance gap lies in the fact that existing RMI designs overlook the importance of data partitioning during model training. Many current RMIs, including the original proposal~\cite{kraska2018case}, typically emphasize on the ``model-fitting'' aspect during index construction. Since the fanout and depth of the index are predetermined (and manually tuned), each leaf node is assigned a corresponding subset of the data and tries to train a model to fit the cumulative distribution function (CDF) of the assigned data as accurately as possible.

The consequence of such a rigid data partitioning strategy contributes to the large performance gap between synthetic and real-world datasets for RMIs. For synthetic datasets, the local data distribution at each leaf node is relatively ``smooth'' so that the expected prediction error is small, leading to a fast last-mile search. Real-world datasets, however, are much ``bumpier'', and the degree of ``bumpiness'' varies across data ranges. Such irregularity in data distribution makes it difficult for simple models (e.g., linear regression) to achieve accurate predictions under predetermined or manually-tuned data partitions. Larger prediction errors require more memory accesses to correct and, therefore, hurt the index performance. Prior work such as~\cite{galakatos2018tree,ding2019alex,wu2021updatable} has attempted to automate partitioning during index construction. Their approaches, however, are heuristic-based and only work on a fixed model type, i.e., linear model.

In this paper, we propose to address the issue by explicitly incorporating data partitioning into the RMI construction process. First, we propose a new cost model that considers the data partitioning aspect for RMI training. The effectiveness of data partitioning is quantified using the entropy~\cite{gray2011entropy} over the number of data points in each partition. The intuition is that larger entropy indicates smaller partition sizes and a more even size distribution. Smaller partitions are preferable because they facilitate leaf-node training and produce more accurate models, especially on non-linear datasets. In addition, smaller child partitions flatten the hierarchical structure of RMI, and thus reduce cache misses. Meanwhile, more even child partitions lead to a more balanced tree structure.

Based on the new cost model, we formalize the index construction problem as an optimization problem and solve it using an algorithm combining the greedy and the dynamic programming approaches. Unlike CDFShop~\cite{marcus2020cdfshop}, where the node types for each layer are determined ahead of time, our algorithm selects the best model and the best partition fanout for each node automatically at construction time without the need for recompilation.

Finally, since memory accesses dominate the lookup performance for RMIs~\cite{sosd-vldb}, we design the memory layout for each node in a cache-aware manner. Specifically, we require the size for every node with a model to be exactly the cacheline size (i.e., 64 bytes) so that each node visit (or model inference) incurs at most one cache miss. For leaf nodes containing data points across multiple cachelines, we adopt a two-level B+ Tree design, consisting of a 64-byte root and multiple 256-byte data blocks. Such a cache-aware layout can effectively reduce the number of memory accesses during the last-mile search as in existing solutions~\cite{marcus2020cdfshop, ding2019alex,radixspline,kraska2018case}, especially on real-world datasets, with only a small space overhead. In addition, fixed-sized nodes facilitate memory prefetching where memory accesses are parallelized to further reduce the access latency.

We present our novel \textbf{C}ache-\textbf{A}ware \textbf{RMI} framework, called CARMI, that implements the new ideas introduced above with six example tree node (model) types. Our experimental study shows that CARMI outperforms all baselines on both our microbenchmark and SOSD benchmark. In our microbenchmark, CARMI achieves an average speedup of 2.2$\times$ (up to 4.2$\times$) and 1.9$\times$ (up to 7.2$\times$) compared to B+ Tree and ALEX, respectively, while only using about 0.77$\times$ memory space of B+ Tree. On the SOSD benchmark, CARMI achieves an average speedup of 2.5$\times$ (up to 3.0$\times$)/1.5$\times$ (up to 2.3$\times$) compared to B+ Tree/ALEX, respectively. Compared to its closest competitor RMI, which has been carefully tuned for each dataset in advance, CARMI is still 1.2$\times$ faster on average (up to 1.5$\times$).

We make the following contributions:
\begin{itemize}[leftmargin=*]
\vspace{-3pt}
\item We identify that the inflexibility of data partitioning for learned indexes is one of the key reasons why there is a large performance gap when applying them to synthetic and real-world datasets.
\item We propose a new cost model for RMI training, which uses entropy across partition sizes to measure the effect of data partitioning on index performance.
\item We formalize the index construction problem as an optimization problem, and propose an algorithm to solve it efficiently and automatically.
\item We propose CARMI, a novel RMI framework implementing the new cost model and automatic node selection algorithm. CARMI also uses a new memory layout that is more cache-friendly, especially for the last-mile search.
\item We conduct a series of experiments to demonstrate the superior performance and robustness of our framework.
\end{itemize}

The remainder of this paper is outlined as follows: In Section~\ref{sec:preliminary}, we review the RMI framework and discuss the two ways of viewing RMI: model fitting vs. data partitioning. In Section~\ref{sec:framework}, we derive a cost model for the entire index structure, and introduce entropy as a metric for characterizing the node performance. Section~\ref{sec:constructionAlg} discusses the cost-based hybrid index construction algorithm, which is used to choose different node settings flexibly to construct the optimal index structure during runtime. In Section~\ref{sec:cache}, we explain the cache-aware designs of CARMI, including the new memory layout and a prefetching mechanism. The experimental setup and results are shown in Section~\ref{sec:experiment}. We discuss the possible extension directions and future works in Section~\ref{sec:discussion}. Finally, we discuss related work in Section~\ref{sec:related} and conclude in Section~\ref{sec:conclusion}. \techreport{Some implementation details, experimental settings and proof of theorem can be found in the appendix.}

\section{Motivation and CARMI}\label{sec:preliminary}
\subsection{RMI and Data Partitioning}\label{sec:partition}
\begin{figure}[ht]
  \centering
  \vspace{-0.4cm}
  \includegraphics[width=\linewidth]{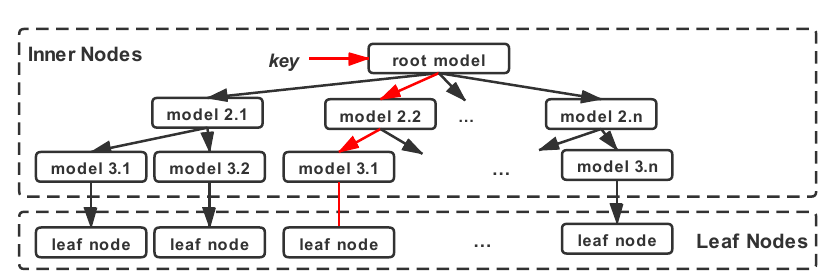}
  \vspace{-0.5cm}
  \caption{Learned Index}
  \Description{Learned Index}
  \label{fig:learnedIndex}
  \vspace{-0.3cm}
\end{figure}

Figure~\ref{fig:learnedIndex} shows the structure of the Recursive Model Index (RMI), an ML-based index framework. Each inner node in the RMI represents an order-preserving regression model: the model $f: k \rightarrow idx$ takes a key $k$ as input, and outputs an integer $idx \in \{1, 2, \ldots, c\}$, where $c$ is the number of child nodes of this inner node. The order-preserving property guarantees $k_1 \leq k_2 \Rightarrow f(k_1) \leq f(k_2)$ so that an RMI can answer range queries correctly. At the bottom layer, leaf nodes are trained using linear models to fit the underlying data points. Searching the RMI given $k$ proceeds as follows: starting from the root, we evaluate the model to determine which child node to visit for the next step. This process is repeated until a leaf node is reached. Finally, we perform a binary search (bounded by the maximum error) to retrieve the matching records.

Existing RMI designs~\cite{kraska2018case, marcus2020cdfshop, ding2019alex, radixspline, stoian2021plex, galakatos2018tree, ferragina2020pgm} view the index construction problem from the perspective of model fitting. In this view, models are trained to minimize a loss function (e.g., squared loss function) to best fit the CDF of a given dataset. Specifically, the root model targets at fitting the CDF of the entire dataset, while the leaf-node models try to fit their local distributions. Under such a problem setting, however, the index fanout and depth are typically predetermined before model training, and thus wasting opportunities to improve model accuracies through a more flexible data partitioning.

Approaching RMI construction via a ``model-fitting'' view can lead to a significant performance drop when shifting from synthetic datasets to real-world datasets. Because real-world datasets are often non-smooth and non-linear, as shown in Section~\ref{sec:sosd}, simple models such as linear regression are not able to fit well in certain CDF ranges if the partitioning is too coarse-grained. Inaccurate predictions from the models then require additional procedures such as large range scans to finish the ``last-mile'' search. These large-range searches span many cachelines and require multiple memory accesses, resulting in performance degradation.

In this paper, we propose to look at RMI construction from a ``data-partitioning'' view, where the models aim to partition a given dataset more evenly into smaller chunks to form a flatter and more balanced tree. Moreover, larger fanouts (i.e., more child partitions) of the upper-level models are preferable in terms of the overall prediction accuracy because lower-level nodes now can work on smaller local datasets to reduce their maximum model prediction errors. The trade-off for more partitions, however, is the space overhead due to more nodes and the performance hit due to potentially more complex models. Therefore, we include the effectiveness of data partitioning in our new cost model along with search time and space, and develop an algorithm to train an RMI automatically using an objective function derived from the cost model.

To quantify the effectiveness of data partitioning, we propose to use entropy, an information-theoretic metric~\cite{gray2011entropy}. Suppose an inner node $M$ distributes a total of $n$ data points into $c$ different child nodes\footnote{Note that the number of child nodes for an inner node is part of the model configurations, and we need to determine its optimal value when constructing the index.}, then the entropy $H(M)$ is defined as: $H(M) = -\sum_{i=1}^{c} p_i \log_2 p_i$, where $p_i = \frac{n_i}{n}$ and $n_i$ is the number of data points allocated to the $i$-th child node. As mentioned in the introduction, larger entropies mean that datasets are divided more evenly into smaller subsets, which is more desirable as discussed above. Further, the entropy also helps establish the notion of local node efficiency, as described in Section~\ref{sec:theoreticalAnalysis}, which combines the time and space cost of a single node and its dataset partitioning utility into a single metric. With this metric, we can effectively compare models locally without global information, thus speeding up index construction.

\subsection{Overview of CARMI}\label{sec:overview}
In this paper, we extend the RMI framework and propose a refined general RMI framework that can automatically build suitable and updatable indexes for given datasets, called Cache-Aware RMI (CARMI). CARMI retains RMI's core idea of replacing traditional indexes with ML models and has similar procedures for querying data points, but differs in specific designs. Specifically, in our new memory layout (\cref{sec:cache}), all tree nodes are limited to 64 bytes and stored in a single array. Then, according to the independent characterization of each tree node (\cref{sec:framework}), CARMI can use different nodes to handle different sub-datasets and link them into an index tree.

Nodes in CARMI are explicitly classified into two categories: inner nodes and leaf nodes. Inner nodes are intermediate bridges between root and leaf nodes, using models to determine the next branch. Leaf nodes are similar in concept to leaf nodes of B+ Tree, but they only store data points conceptually by storing pointers to data blocks containing actual data points.

For a given dataset, CARMI uses a hybrid construction algorithm (\cref{sec:constructionAlg}) to solve the index construction problem, whose optimization objective is to minimize the weighted sum of time and space costs (\cref{sec:framework}). With this algorithm, CARMI can automatically construct indexes with good performance at runtime.

\begin{figure}[ht]
  \vspace{-0.3cm}
  \centering
  \includegraphics[width=\linewidth]{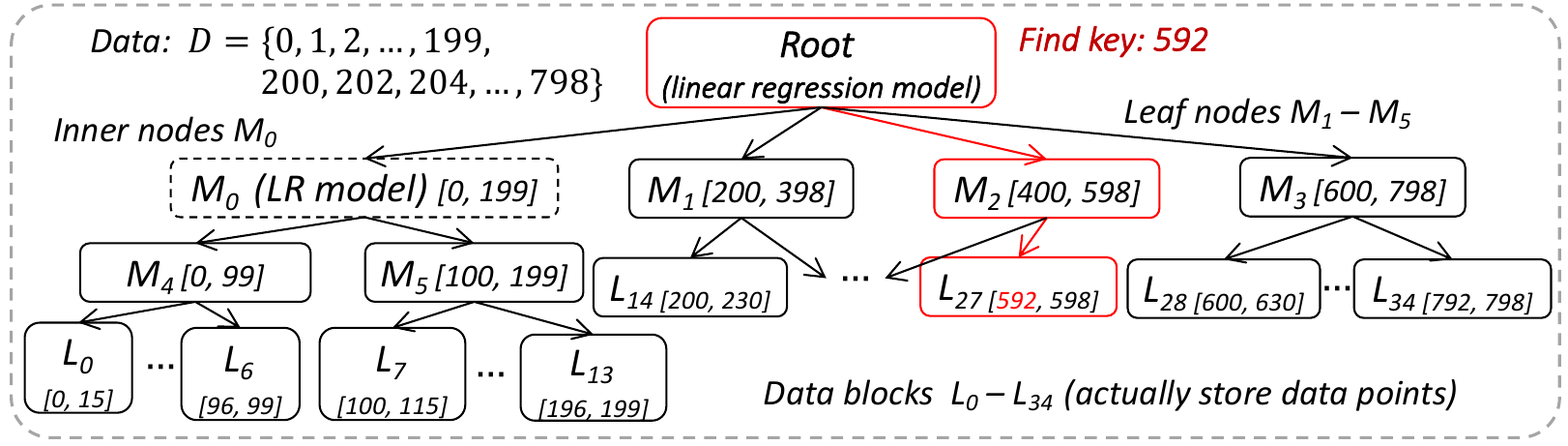}
  \vspace{-0.4cm}
  \caption{A Simple Example of CARMI}
  \Description{Example}
  \label{fig:example}
\vspace{-0.3cm}
\end{figure}

Then, we illustrate the design of CARMI via a concrete example:

\begin{example}\label{example1}
Consider a dataset in which there are 500 data points $D = \{0, 1, \cdots 199, 200, 202, \cdots, 798\}$. Then CARMI can be a four-layer structure, as shown in Figure~\ref{fig:example}. The top layer is the root node with a linear regression (LR) model ($idx = \lfloor 0.005 \times key \rfloor$), which determines the index of the next node for each given key. The first 200 data points are managed by an LR inner node and its two child leaf nodes. The remaining 300 data points are directly managed by 3 leaf nodes ($M_1-M_3$). Each leaf node is linked to 7 data blocks at the bottom and uses them to store data points.

Suppose that we need to access the record with key value $592$. We first access the root node and use its model to calculate the index of the next node ($idx = \lfloor 0.005 \times 592 \rfloor = 2$). After obtaining the content of $M_2$, we use its strategy to get the index of the data block ($L_{27}$), and finally, search for the key value $592$ within $L_{27}$. Since the root node is frequently accessed, we assume that it is always in the cache memory. Therefore, for this example, we only need two random memory accesses: the leaf node $M_2$ and the data block $L_{27}$, respectively.
\end{example}

\section{Cost Model of CARMI}\label{sec:framework}
In this section, we describe the new cost model of CARMI and its application in the index construction problem in detail. Specifically, to quantify the performance of tree nodes in a standalone fashion and lay a solid foundation for index construction, we first characterize each tree node from three dimensions: time cost, space cost, and entropy, and use them as building blocks to derive a cost model for the entire index structure. Since the analysis is performed from a generic perspective, the CARMI framework can accommodate any new types of tree nodes as long as they can be represented accordingly. Then, we use the cost model as an optimization objective to find the most suitable node design and tree structure at runtime. Some example tree node designs are briefly described in this paper, together with their performance characterization. These designs are all included in our open-source implementation of CARMI~\cite{carmiGithub}, and used in our experimental study in Section~\ref{sec:experiment}.

\begin{figure*}[ht]
  \centering
  \vspace{-0.7cm}
  \includegraphics[width=\linewidth]{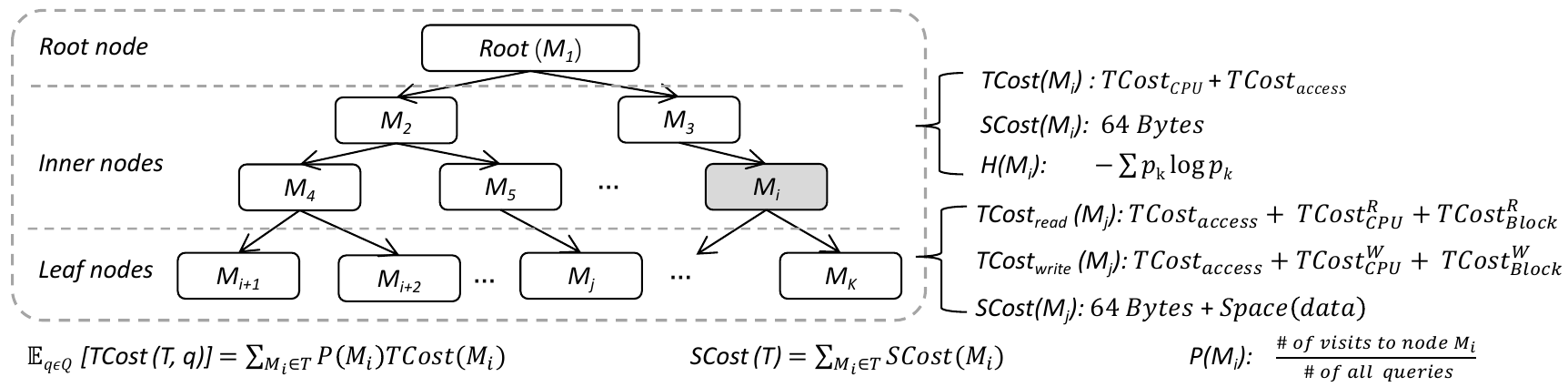}
  \vspace{-0.4cm}
  \caption{Cost Model of CARMI}
  \Description{Cost model}
  \label{fig:costModel}
\end{figure*}

The rest of the section is organized as follows. We analyze the inner/leaf nodes from three perspectives in Section~\ref{sec:innerNodes} and~\ref{sec:leafNodes}, respectively. The cost model of the entire index and the formal formulation of the index construction problem is described in Section~\ref{sec:optProblem}. Finally, in Section~\ref{sec:specific}, we briefly talk about a few specific tree node designs that we have implemented.

\subsection{Inner Nodes} \label{sec:innerNodes}
The main functionality of inner nodes is to determine which branch to go through, so that we can quickly map a given key to its corresponding leaf node. In the following, we discuss three separate dimensions for characterizing inner nodes: the time required for predicting the next branch, the space cost of the node, and the degree of uniformity of data points partitioning.

\subsubsection{Time}
The time cost of an inner node includes two parts: the access time and the computation time, denoted as $TCost_{access}$ and $TCost_{CPU}$, respectively. $TCost_{access}$ refers to the time to read the node content, which is equal to the latency of the main memory due to our cache-aware design in Section~\ref{sec:cache}. $TCost_{CPU}$ is the time required for a model to compute the index of the next node, which only depends on the model type. Then, the total time required for an inner node $M$ to predict the next branch is: $TCost(M) = TCost_{CPU} + TCost_{access}$.

\subsubsection{Space}
The space cost of an inner node $M$ is the total amount of space in bytes, which is denoted as $SCost(M)$.

\subsubsection{Entropy}\label{sec:entropy}
We use the entropy metric to characterize the ability of an inner node to partition a dataset evenly. The entropy of an inner node $M$ is: $H(M) = -\sum_{i=1}^{c} p_i \log_2 p_i$ (details in Section~\ref{sec:partition}).

\subsubsection{Root Node}
In CARMI, the root node is handled differently compared to inner nodes: since the root node is always accessed during lookup procedures, we can assume that it is always available in the cache memory. Because of this, $TCost_{access}$ of the root node is equal to the latency of cache memory.

\subsection{Leaf Nodes} \label{sec:leafNodes}
Leaf nodes are used to manage the actual data points and can also be characterized in terms of time and space cost. The third dimension is the capacity for storing data points (i.e., how many data points can be stored in the leaf node).

In CARMI, we design a new type of leaf node similar to a two-level B+ tree node. Its root is 64 bytes and contains pointers to multiple 256-byte data blocks that store data points. Then, three dimensions of leaf nodes are described as follows.

\subsubsection{Time}
For leaf nodes, the time cost of two specific operations is analyzed: insert a new data point and lookup a data point\footnote{Deletion and update operations are not discussed since they are similar to the read access operations (we adopt a lazy deletion approach).}.

Similar to inner nodes, the time to access the node ($TCost_{access}$) is equal to the main memory access latency. Due to the new leaf node, finding a data point requires first finding the index of the data block, and then locating the data point within the block. Their time costs are denoted as $TCost_{CPU}^{R}$ and $TCost_{Block}^{R}$, respectively.

As for the insert operation, the content of both leaf node and data blocks might need to be changed accordingly. To reflect this, the time cost is denoted using a different superscript ($TCost_{CPU}^{W}$ and $TCost_{Block}^{W}$).

Overall, the time cost of a leaf node is modeled as:
\begin{equation}\label{eq:timecost_leaf}
	\begin{split}
	  \textstyle
	 &TCost_{read}(M) = TCost_{access}+TCost_{CPU}^{R}+TCost_{Block}^{R}\\
	 & TCost_{write}(M) = TCost_{access}+TCost_{CPU}^{W}+TCost_{Block}^{W}
	\end{split}
\end{equation}

\subsubsection{Space}
The space cost of a leaf node $M$, consists of the bytes of metadata and data blocks. Then the total space cost is:
\begin{equation}\label{eq:spacecost_leaf}
	\begin{split}
	  \textstyle
	 SCost(M) = SCost_{leaf} + Space(data)
	\end{split}
\end{equation}
where $SCost_{leaf}$ is 64 bytes, and $Space(data)$ is the total amount of space occupied by the data blocks.

\subsubsection{Capacity of Leaf Nodes}\label{sec:capacity}
The capacity of leaf nodes refers to the total number of data points stored in a leaf node, and it depends on both the total amount of space allocated for data points as well as the way these data points are arranged. For example, if we need to make room for subsequently inserted data points to reduce the latency of the insert operation, then the capacity of the leaf node will be reduced accordingly.

\subsection{The Optimization Problem} \label{sec:optProblem}
Based on the above analysis, we can define a cost model for the entire index structure. The time cost of queries can be estimated by utilizing the above analysis results: For any query $q$ and index structure $T$, let the traversal path of $q$ in $T$ be $M_1(root) \rightarrow M_2 \rightarrow \ldots \rightarrow M_k(leaf)$, then the time cost of $q$ can be approximated as:
\begin{equation}
  TCost(T, q) = \sum_{i=1}^k TCost(M_i)
\end{equation}

The space cost of the index structure is simply the sum of the space cost of all inner nodes and leaf nodes:
\begin{equation}
  SCost(T) = \sum_{M_i \in T } SCost(M_i)
\end{equation}

Now we can formalize the problem of index construction as an optimization problem: We would like to find the optimal index structure that minimizes the average time cost of each data lookup operation, under a certain space cost budget. Here the average time cost is evaluated with respect to a fixed query workload known in advance. The query workload can be constructed using recent history queries from users. In most cases, recent history queries will faithfully reflect the characteristics of queries we expect to see in the future. If there are no history queries available, then we can use a uniform access workload instead, in which each data point is accessed exactly once.

The problem of finding an optimal tree structure is formulated as follows: 
\begin{problem}\label{prob:m_budget}
	Let $Q = \{q_1, \ldots, q_m\}$ be a collection of queries, and $D = \{d_1, \ldots, d_n\}$ be the collection of keys to be maintained in the index structure. Find the optimal index tree structure $T$ such that $\mathbbm{E}_{q \in Q}[TCost(T, q)]$ is minimized, under the constraint that the total space cost of $T$ does not exceed a fixed budget $B$: $SCost(T) \leq B$.
\end{problem}

Problem~\ref{prob:m_budget} is a constrained optimization problem, and a widely used technique for solving such problems is to use the Lagrange multiplier method~\cite{beavis1990optimisation} to transform it into an unconstrained problem. By utilizing this technique, Problem~\ref{prob:m_budget} can be transformed into a roughly equivalent form with a linear combination of time and space cost as objective, which is relatively easier to optimize:
\begin{problem}\label{prob:linear}
	Let $Q = \{q_1, \ldots, q_m\}$ be a collection of queries, $D = \{d_1, \ldots, d_n\}$ be the collection of keys to be maintained in the index structure, and $\lambda$ be a positive constant parameter. Find the optimal index tree structure $T$ such that $\mathbbm{E}_{q \in Q} [TCost(T, q)] + \lambda SCost(T)$ is minimized.
\end{problem}

Problem~\ref{prob:linear} suggests that we ultimately want to minimize a weighted sum of time and space cost of the index, and we will describe an algorithm for solving it in Section~\ref{sec:constructionAlg}.

Let $P(M_i)$ be the fraction of history queries passing through a tree node $M_i$. Then the expression $\mathbbm{E}_{q \in Q}[TCost(T, q)]$ can be rearranged into an alternative form, which is sometimes more convenient:
\paper{ \vspace{-0.2cm}}
\begin{equation}\label{eq:eq}
  \mathbbm{E}_{q \in Q} [TCost(T, q)] = \sum_{M_i \in T} P(M_i)TCost(M_i)
\end{equation}

Finally, a summarization of CARMI cost model can be found in Figure~\ref{fig:costModel} for fast reference.

\subsubsection{Theoretical Analysis}\label{sec:theoreticalAnalysis}
In the following analysis, we assume the history queries to be a uniform access of the data points for simplicity, in such a case the value of $P(M_i)$ is the same as the total fraction of data points in the subtree of $M_i$. With this assumption, the following theorem establishes a connection between $\sum_{i} P(M_i) H(M_i)$ and the total number of data points $n$.

\begin{theorem}\label{theo:entropy}  
Let $T = \{M_1, \ldots, M_K\}$ be an index structure with $K$ nodes in total. $P(M_i)$ represents the ratio of data points in node $M_i$ relative to the total number $n$. Then we have: $\sum_{i=1}^K P(M_i)H(M_i) = \log_2n$, where for leaf nodes, $H(M_i)$ is defined as $\log_2(Capacity(M_i))$ and $Capacity(M_i)$ is the capacity of leaf nodes (\cref{sec:capacity}).
\end{theorem}

\paper{The proof of this theorem can be found in our technical report~\cite{zhang2021carmi}.} \techreport{The proof of this theorem can be found in the appendix.} Essentially, Theorem~\ref{theo:entropy} states that the weighted sum of the entropy of all tree nodes is always a constant value. In other words, entropy characterizes the ``contribution" of each tree node to the index structure: the higher entropy each tree node contributes, the less overall number of tree nodes we need in the index structure.

It is also of interest to compare Theorem~\ref{theo:entropy} with the optimization objective of Problem~\ref{prob:linear}. We can rewrite the objective using Equation~\ref{eq:eq}:
\paper{
    \vspace{-0.4cm}}
\begin{equation*}\label{eq:rewrite}
  \text{Objective} = \sum_{i=1}^K [P(M_i)TCost(M_i) + \lambda SCost(M_i)]
\end{equation*}

Comparing with Theorem~\ref{theo:entropy}, we see that intuitively each tree node $M$ contributes $P(M)H(M)$ entropy-wise to the index structure while incurring $P(M)TCost(M) + \lambda SCost(M)$ cost to the overall objective. Thus the ratio of these two terms can be used to quantify the local efficiency of node $M$:
\begin{equation} \label{eq:greedy}
	\begin{split}
    \text{cost-ratio(M)} & = \frac{P(M) TCost(M) + \lambda(SCost(M))}{P(M)H(M)}\\
  \end{split}
\end{equation}

Generally, we want to minimize the cost-ratio of all tree nodes, especially the ones with a large value of $P(M)H(M)$. For example, if all tree nodes have a cost-ratio less than $c$, then the optimization objective would be bounded by $c\log_2 n$.
 \begin{table}[ht]
 \paper{ \vspace{-0.3cm}}
  \small
  \center
  \caption{The Mechanism of Various Types of Nodes}
  \label{tab:nodes}
  \vspace{-0.2cm}
  \begin{tabular}{m{0.15\linewidth}m{0.76\linewidth}}
    \toprule
    Node & Mechanism\\
    \midrule
    \midrule
    LR & Use a LR model to determine the corresponding branch\\
    \midrule
    P. LR & Similar to LR nodes, but use a piecewise linear regression model instead\\
    \midrule
    Hist & Determine the branch through a histogram lookup table\\
    \midrule
    BS & Use binary search to determine the branch, similar to B+ Tree node\\
    \midrule
    \midrule
    CF Array & A two-layer cache-friendly structure similar to a B+ Tree\\
    \midrule
    Ext. Array & Only store meta data, and data points are stored in an external location\\
    \bottomrule
  \end{tabular}
  \paper{ \vspace{-0.5cm}}
 \end{table} 

\subsection{Specific Implementation} \label{sec:specific}
We have implemented four types of inner nodes and two types of leaf nodes in CARMI. For inner nodes, they use either linear regression, piecewise linear regression, binary search or histogram models to predict the next branch. For leaf nodes, we implemented two different structures: cache-friendly array (CF array) and external array. CF array leaf nodes store data points compactly in the data blocks in a sequential manner, and leaf node itself stores the minimum key values of data blocks. External array leaf nodes are used for primary index structures, where the original data points are already sorted according to the key value and stored in an external location. In such a case, we only need to store pointers to external locations in the leaf node.

\techreport{
    \begin{figure*}[ht]
    \centering
    \vspace{-0.9cm}
    \includegraphics[width=\linewidth]{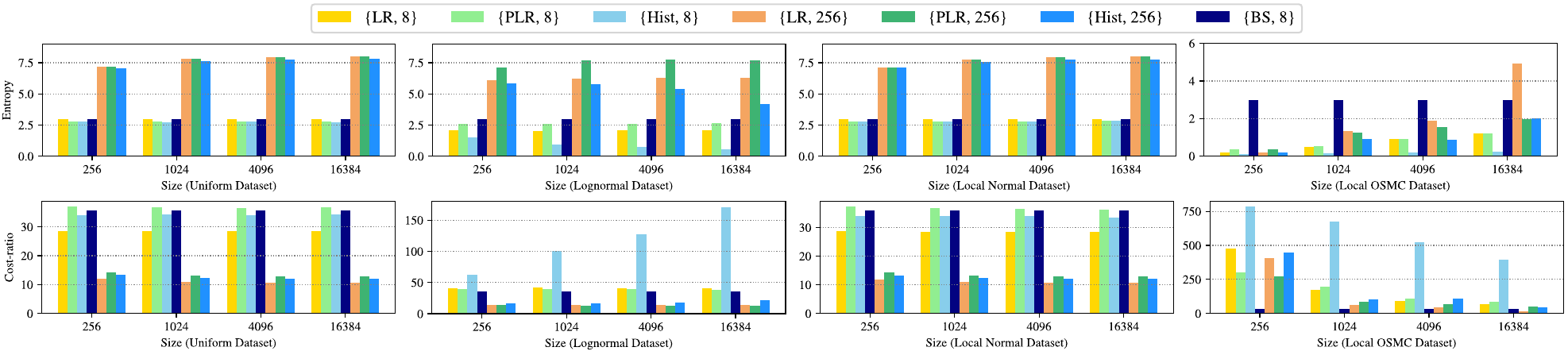}
    \vspace{-0.5cm}
    \caption{Entropy and cost-ratio, where each case is a node setting consisting of node type and number of children.}
    \Description{Entropy on typical datasets, where each bar is a node setting, including the node type and the number of child nodes.}
    \label{fig:entropyData}
  \end{figure*}
}

Note that the choice of inner/leaf nodes can be flexibly determined at runtime and do not need to agree on a single one throughout the tree structure. Table~\ref{tab:nodes} summarizes the mechanism and characteristics of each tree node type. \paper{The specific implementation details can be found in our technical report~\cite{zhang2021carmi}, along with the empirical performance, entropy, and local efficiency of these nodes.} \techreport{The specific implementation details can be found in the appendix.}

\techreport{
\subsubsection{Empirical Performance}
In order to evaluate the empirical performance of tree nodes, we have implemented a profiler program to evaluate the CPU computation time required for each node to obtain the position of the next branch and the time to obtain their contents from memory. Table~\ref{tab:dimensions} shows the empirical performance of these node types based on our evaluation on an Ubuntu platform equipped with an AMD Ryzen 3700X 8-Core Processor and 32GB RAM.
\begin{table}[ht]
  \vspace{-0.2cm}
  \small
  \center
  \caption{The Empirical Values of Nodes in CARMI}
  \label{tab:dimensions}
  \vspace{-0.2cm}
  \begin{tabular}{ccccc}
    \toprule
    \multirow{2}*{Type} &\multirow{2}*{Node}&  Space & \multicolumn{2}{c}{Time (ns)}\\
    ~&~& (Bytes) & Access &CPU\\
    \midrule
    \multirow{2}*{Root} &LR&20  & 8.29& 3.25\\
    ~&P. LR&76  &11.24 & 18.38\\
    \midrule
    \multirow{4}*{Inner} &LR&\multirow{4}*{64} & \multirow{4}*{80.09}&5.2\\
    ~&P. LR&~ &~ &22.8\\
    ~&Hist&~ & ~ &14.1\\
    ~&BS&~& ~ &27.3\\
    \midrule
    \multirow{3}*{Leaf} &CF (leaf node) &64 &80.09 & 25.4 \\
    ~&CF (data block) &  256& 90.09& 53.8\\
    ~&Ext. Array&64&80.09&117.1$-$328.3\\ 
    \bottomrule
  \end{tabular}
 \paper{ \vspace{-0.3cm}}
\end{table}

We also evaluate the entropy and the cost-ratio of inner nodes on three synthetic datasets and one real-world dataset in Figure~\ref{fig:entropyData}. Detailed descriptions of the datasets can be found in Section~\ref{sec:setup}. We provide the results of two different cases: (a) on two full synthetic datasets: uniform and lognormal datasets, (b) on random consecutive $n$ data points in normal and OSMC datasets to show the results of local distributions. The number of key values in each case is 256, 1024, 4096, 16384, and the number of child nodes is 8 and 256, respectively.
  
\textbf{\textit{1) Analysis of Root Nodes.}} Since the root node is frequently accessed and always in the cache memory, we do not need to limit the size of the root node, and can choose the node with a larger entropy to improve the overall performance. The LR node only needs a small amount of time (3.25 ns) to calculate the position of the next branch, while the P. LR node needs 18.38 ns.

\textbf{\textit{2) Analysis of Inner Nodes.}} Due to the cache-aware design, the space cost of inner nodes is the same, but the time cost is different. The LR node enjoys the minimum CPU time cost of 5.2 ns, while other nodes need more time to get the position of the next branch. Note that even for the most costly BS nodes (27.3 ns), the CPU computation time is still much less than the access time, confirming our earlier intuition that we should try to minimize the number of memory accesses.

The utility of inner nodes on different datasets can be effectively reflected by entropy, and can be combined with time/space cost to help flexibly determine the appropriate nodes during construction. As shown in Figure~\ref{fig:entropyData}, LR nodes handle linear datasets well (e.g., uniform dataset, local normal dataset), obtaining the largest entropy close to $\log_2 c$ and lowest cost-ratio. However, ML nodes cannot achieve good utility with the same number of children as BS nodes since the distribution of local OSMC dataset is  highly non-linear. In this case, BS nodes that can maintain the entropy of $\log_2 c$ are most suitable. Overall, these results validate that the data partitioning view is valuable and that it makes sense to use different nodes to build indexes flexibly.

\textbf{\textit{3) Analysis of Leaf Nodes.}} The time cost of these two leaf nodes suggests that they are suitable for different situations. Since LR models and binary searches are required to lookup data points, the time cost of external leaf nodes varies greatly (117.1-328.3 ns) depending on the capacity and data distribution. When search ranges are small, this node only takes 197.19 ns to find the exact location, and thus, it is more suitable for linear datasets (e.g., YCSB dataset). CF leaf nodes can be widely used on various datasets, since only two memory accesses are required to obtain data points regardless of data distributions.
}

\section{Index Construction Algorithm}\label{sec:constructionAlg}
In Section~\ref{sec:optProblem}, we have shown that the optimal index tree structure can be constructed by minimizing the weighted sum of the time and space cost of the index structure (see Problem~\ref{prob:linear}). In this section, we describe an algorithm for solving it. 

First, let us rearrange the optimization objective as follows: 
\begin{align} \label{eq:construction}
		\mathbbm{E}_{q \in Q} &[TCost(T, q)] + \lambda SCost(T) = TCost(root) + \lambda SCost(root)\notag\\
		&+ \sum_{i=1}^c [\frac{\left| Q_{T_i} \right|}{\left| Q \right|}\mathbbm{E}_{q \in Q_{T_i}} [TCost(T_i, q)] + \lambda SCost(T_i)]
\end{align}
where $c$ represents the number of child nodes of the root node, $T_i$ is the sub-index tree of the $i$-th child node, and $Q_{T_i}$ is the collection of queries that access $T_i$. Note that the number of child nodes is chosen from an exponentially increasing sequence to reduce training time, such as the powers of 2.

As we can see, the terms inside the square brackets have a very similar form compared to the original objective. In other words, once the root node of this subtree is fixed, the original optimization problem can be broken down into several independent sub-problems of similar form, and each sub-problem can be solved independently. For each sub-problem, we have two different algorithms for solving it:

\begin{itemize}[leftmargin=*] 
	\item \textbf{Node selection algorithm:} A greedy algorithm that only considers local information to construct nodes, which is used when the subtree manages a large dataset (e.g., the root node).
	\item \textbf{Dynamic programming (DP) algorithm:} This algorithm is guaranteed to find the optimal sub-structure but is slower, and we only use it when the sub-dataset is small.
\end{itemize}

These two algorithms are used in combination in our construction algorithm. Then, the overall workflow for constructing the index structure for a given dataset is: 

\begin{itemize}[leftmargin=*]
	\item First, we choose a root node setting using the greedy node selection algorithm.
	\item Next, we use the setting obtained in the previous step to assign the dataset to the child nodes of the root, and then construct a sub-index tree over each assigned sub-dataset.
	\item Each child node chooses the appropriate algorithm to build the sub-index structure according to the size of sub-dataset.
	\item Finally, all the sub-structures are merged together and linked to the root node to form the complete index tree.
\end{itemize}

The rest of this section is organized as follows: The greedy node selection step is described in Section~\ref{sec:greedy} and the DP algorithm is described in Section~\ref{sec:dpAlg}. \techreport{We provide a simple example to illustrate our construction algorithm in Section~\ref{sec:constructionExample}.}

\subsection{Node Selection Algorithm} \label{sec:greedy}
For nodes that manage a large fraction of the dataset, such as the root node, we hope to quickly select a suitable node design using only local information. Below, we develop a greedy node selection algorithm to find the locally optimal solution without considering the design of the lower-level nodes:

\begin{itemize}[leftmargin=*]
	\item For the current node, the node selection algorithm enumerates various possible node types and considers a number of choices for the number of child nodes. For each enumerated setting, we calculate the time cost, space cost and entropy of the node. 
	\item Then, we calculate the cost ratio of each setting using Equation~\ref{eq:greedy} in Section~\ref{sec:optProblem}, and select the setting with the minimum cost ratio for the current node.
	\item Using the setting (model and child number) obtained in the previous step, we assign each data point to one of the child nodes of the current node.
	\item Finally, the sub-index tree of each child node is constructed recursively using either this algorithm or the dynamic programming algorithm in the next section.
\end{itemize}

\techreport{
	\begin{algorithm}[ht]
		\caption{The Greedy Node Selection Procedure}
		\label{alg:nodeSelectionAlg}
		\begin{algorithmic}[1]
			\REQUIRE ~~ keys $K[]$, access frequency $f[]$, query number $m$;
			\ENSURE ~~ local optimal design $M$
			\STATE $OptimalValue \leftarrow \infty$
			\STATE $P \leftarrow \sum_{j=1}^{\left|f\right|}  f_j / m$
			\FOR{ every inner node design $M_i$}
				\STATE $K_1, \ldots, K_{c_i} \leftarrow \text{Partition}(K, M_i)$
				\STATE $ChildSCost \leftarrow c_i \times 64$ bytes
				\STATE $Cost \leftarrow TCost(M_i) + \lambda \times ChildSCost/ P$
				\FOR{$j = 1$ \TO $c_i$}
					\STATE $p_j \leftarrow  \left|K_j\right| / \left|K\right| $
				\ENDFOR
				\STATE $H(M_i) \leftarrow - \sum_{j=1}^{c_i} p_j \log_2 p_j$
				\IF{$(Cost / H(M_i)) < OptimalValue$}
					\STATE $OptimalValue \leftarrow Cost / H(M_i)$
					\STATE $M \leftarrow M_i$
				\ENDIF
			\ENDFOR
			\RETURN $M$
	\end{algorithmic}
	\end{algorithm}

The pseudocode of the greedy node selection algorithm is shown in Algorithm~\ref{alg:nodeSelectionAlg}. Note that we are not directly using the space cost of the node itself for computing the cost-ratio, but rather the total space cost of the child nodes of node $M$. The reason behind such a design is that the original cost-ratio would always favor large number of child nodes (since the time/space cost of the node itself is fixed). This change allows the greedy algorithm to choose a more reasonable node setting.
}

\subsection{Dynamic Programming Algorithm}\label{sec:dpAlg}

As shown in Equation~\ref{eq:construction}, the original optimization problem can be decomposed into similar sub-tasks once the root node is fixed. Based on this intuition, we can use a dynamic programming algorithm to solve this problem:
\begin{itemize}[leftmargin=*]
	\item \textbf{State:} The state of each step is the dataset handled by the current node, denoded as $D_{l,r}$, where $l$ and $r$ are the left and right index of the entire dataset. Let $Q_{l,r}$ be the collection of queries that access $D_{l,r}$, and $T_{l,r}$ be the sub-index tree for $D_{l,r}$. Then the corresponding entry $cost[l,r]$ in DP table is the value of the following expression:
	\vspace{-3pt}
	\begin{equation} \label{eq:DPstate}
		cost[l, r] = \min_{T_{l,r}} \{ \frac{\left| Q_{l,r} \right|}{\left| Q \right|}\mathbbm{E}_{q \in Q_{l,r}} [TCost(T_{l,r}, q)] +\lambda SCost(T_{l,r})\}
	\end{equation}
	\item \textbf{State transition equation:} Let $S$ be the collection of all possible settings for the root node of sub-index tree, including the type of the root node and the number of the child nodes. Then according to Equation~\ref{eq:construction} and~\ref{eq:DPstate}, the state transition equation is:
	\vspace{-3pt}
	\begin{equation*} \label{eq:dp}
		cost[l, r] = \min_{(M, c) \in S} \{\frac{\left| Q_{l,r} \right|}{\left| Q \right|} TCost(M)+\lambda SCost(M) + \sum_{j=1}^{c} cost[l_j, r_j]\}
	\end{equation*}
	where $M$ and $c$ are the node type and the number of child nodes (if the node is a leaf node, $c$ is 0), respectively, $l_j$ and $r_j$ are the left and right index of the sub-dataset that belongs to the child node $j$. If the dataset is smaller than a certain threshold, then we only consider the leaf node settings in the state transition equation.
\end{itemize}

Specifically, given a sub-dataset $D_{l,r}$, we need to enumerate all its possible settings for the root node, and compute the cost of each setting separately. Then, the setting with minimum cost is selected to construct the current node, and the cost is stored to the entry $cost[l, r]$. For each setting, we first look up the time/space cost of the root node from our cost model, and use the sub-dataset $D_{l,r}$ to train the root node model $M$.  Then, we use the trained model $M$ to assign the dataset to $c$ child nodes and obtain their cost via a memorized search approach: for each subtask of computing $cost[l_{j}, r_{j}]$, the algorithm first checks whether it has been solved before. Then according to the check result, it either returns the minimum cost directly from the DP table or recursively calls the process of the DP algorithm. When the number of data points in the subtree is less than a pre-specified threshold, the algorithm will directly construct a leaf node as the current node. Otherwise, the algorithm considers two cases (leaf nodes or subtrees with inner nodes) and chooses the optimal design according to the transition equation. \paper{The pseudocode can be found in our technical report~\cite{zhang2021carmi}, as well as a simple example to illustrate our construction algorithm.}\techreport{The pseudocode of the DP algorithm is shown in Algorithm~\ref{alg:dp}.}
\techreport{
	\begin{algorithm}[ht]
		\caption{DP Algorithm}
		\label{alg:dp}
		\begin{algorithmic}[1]
			\REQUIRE ~~keys $K[]$, queries $Q[]$, left index $l$, right index $r$;
			\ENSURE ~~optimal structure $T_{l,r}$, $\frac{\left| Q_{l,r} \right|}{\left| Q \right|}\mathbbm{E}_{q \in Q_{l,r}} [TCost(T_{l,r}, q)] + \lambda(SCost(T_{l,r}))$
			\IF{sub-problem $(l,r)$ have been solved before}
			  \RETURN $(T_{l,r}, cost[l,r])$
			\ENDIF
			\STATE /*$kLeafMaxCapacity$ must be larger than $kLeafThreshold$*/
			\IF {$r-l+1 <= kLeafMaxCapacity$}
			\STATE Construct a leaf node $T_{l,r}$ from $K[l, \ldots, r]$ and $Q[l, \ldots, r]$.
			\STATE $OptValue \leftarrow \frac{\left| Q_{l,r} \right|}{\left| Q \right|} \mathbbm{E}_{q \in Q_{l,r}} [TCost(T_{l,r}, q)] + \lambda(SCost(T_{l,r}))$
			\ENDIF
			\IF {$r-l+1 > kLeafThreshold$}
			\FOR{every inner node design $M$}
			  \STATE $(\{l_1,r_1\}, \ldots, \{l_c,r_c\}) \leftarrow \text{Partition}(K[l, \ldots, r], M)$
			  \FOR{$j = 1$ \TO $c$}
				\STATE $(T_{l_j,r_j}, cost[l_j,r_j]) \leftarrow \text{DP}(K, Q,l_j,r_j)$
			  \ENDFOR
			  \STATE $RootCost \leftarrow \frac{\left| Q_{l,r} \right|}{\left| Q \right|} TCost(M) + \lambda (SCost(M))$
			  \IF {$\sum_j cost[l_j,r_j] + RootCost < OptValue$}
				\STATE $OptValue \leftarrow \sum_j cost[l_j,r_j] + RootCost$
				\STATE Construct an index structure $\{T_{l_j,r_j}\}$ from $T_j$ and $M$
			  \ENDIF
			\ENDFOR
			\ENDIF
			\RETURN $(T_{l,r}, OptValue)$
	\end{algorithmic}
	\end{algorithm}
}

Although in principle, the dynamic programming algorithm can be used to construct the optimal index structure for the entire dataset, in practice, it is too slow to handle large datasets. Therefore, we only use it to solve sub-problem that are small enough.

\techreport{
\subsection{A Simple Example}\label{sec:constructionExample}
Here we use the same example as before to demonstrate the entire index construction process.
\begin{figure}[h]
	\vspace{-0.2cm}
	\centering
	\includegraphics[width=\linewidth]{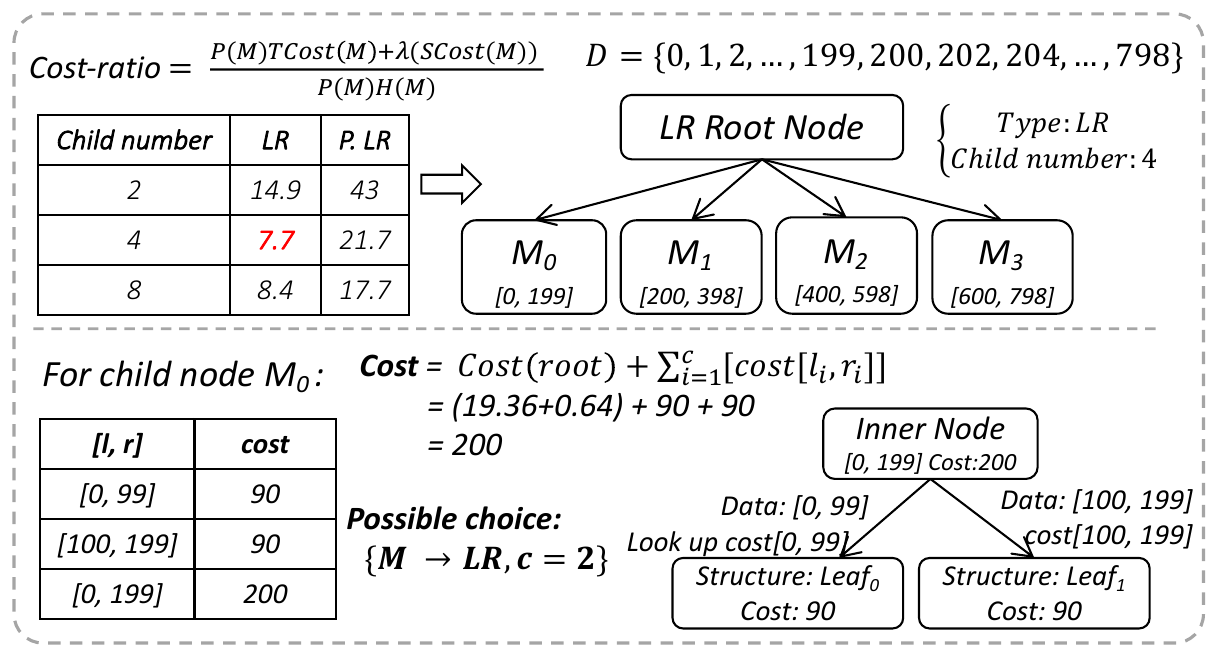}
	\vspace{-0.4cm}
	\caption{Construct the Root/Inner Node}
	\Description{Construct the Root Node}
	\label{fig:buildExample}
\vspace{-0.5cm}
\end{figure}
\begin{example}
We demonstrate our construction algorithm using the dataset $D = \{0, 1, \cdots 199, 200, 202, \cdots, 798\}$ in Figure~\ref{fig:buildExample}. First, we use the greedy node selection algorithm to choose the setting of the root node. We consider two node types: LR and P.LR, and compute their cost ratios with 2, 4, and 8 child nodes, respectively. Among them, the LR root node with 4 child nodes has the minimum cost ratio of 7.7. Therefore, we use it to construct the root node and assign the dataset to the 4 child nodes according to the trained LR model.

We use sub-datasets to construct the optimal sub-trees for 4 child nodes according to the DP algorithm. We need to enumerate all possible settings for them. For example, for node $M_0$ with $D_{0,199}$, a possible setting is that the type is the LR inner node and the number of child nodes is 2. After looking up its cost of 20, we use the trained LR node to assign the dataset to the child nodes and obtain two smaller sub-datasets: $D_{0,99}$ and $D_{100,199}$. According to the state transition equation, the algorithm needs to compute the minimum cost for the two sub-datasets. Here we assume that their cost has already been computed previously, therefore, we only need to look up $cost[0, 99]$ and $cost[100, 199]$ in the DP table (both are 90). Then, we can get the overall cost of 200 by adding up all three costs. After enumerating each possible choice, we use the setting with the minimum cost to construct $M_0$. Then, for $M_1-M_3$, we repeat the above steps and select a leaf node as their root node. Finally, the optimal index structure can be obtained by linking these four subtrees with the root node.
\end{example}
}

\section{Cache-Aware Design}\label{sec:cache}
In this section, we describe the cache-aware design of CARMI with a new memory layout in detail, use some examples to explain the intuition and benefits of such a design, and also discuss the potential opportunities for using memory prefetching instructions to further speed up queries.

The rest of this section is organized as follows: Section~\ref{sec:cache-aware} discusses the details of cache-aware design, and Section~\ref{sec:memoryLayout} describes our new memory layout and the basic lookup and insert operations. The use of memory prefetch will be discussed in Section~\ref{sec:prefetch}.

\subsection{Details of Cache-Aware Design}\label{sec:cache-aware}

B+ Trees with tree nodes occupying exactly a cache line size can outperform standard binary search trees~\cite{rao2000making}. Specifically, each memory access always copies a fixed-size memory content (i.e., cache line size, usually 64 bytes), then indexes that utilize the cache and do not ``waste" any data retrieved into cache memory can generally outperform standard data structure by a large margin.


In CARMI, we employ the same design decision as in B+ Tree and enforce all tree nodes to have a fixed size of 64 bytes. To understand the intuition behind such a design,
we briefly analyze
the performance bottleneck of the lookup procedure in RMI. For an index structure within the RMI framework (shown in Figure~\ref{fig:learnedIndex}), the procedure for accessing a data point is as follows:
\begin{itemize}[leftmargin=*]
  \vspace{-2pt}
  \item We first access the content of the root node, and then use its model to choose a child node of the root.
  \item Subsequently, in each layer, we visit a node chosen by the node in the previous layer, and then use the model of this node to choose one of its child nodes. We repeat this step until we have reached a leaf node.
\end{itemize}

In the above procedure, for each tree node, we need at least one memory access to get its content before performing the model computation. For simple models (e.g., linear models), the computation time is usually less than 20 ns, while a memory access takes about 70-100 ns if the node content is not cached. Therefore, the time required for memory access would actually take up most of the time we spent on data lookup. In other words, the performance bottleneck of RMI is memory access. 

Based on the above analysis, we hope to minimize the number of memory accesses and fully utilize each access during data lookup. In CARMI, we achieve them from the following two aspects.

First, we enforce each node to have a size of exactly 64 bytes to align the cache line. This design allows the node to be fetched with only one memory access. Furthermore, the large node size allows it to store rich information, which helps to reduce the average tree depth, thus reducing the number of needed memory accesses.

To further reduce the number of memory accesses in the last mile, we design a new leaf node called Cache-Friendly array leaf node, which is conceptually similar to a two-layer B+ Tree node. The first layer is a 64-byte root that stores pointers to several 256-byte data blocks in the second layer and the minimum key values of each block. With such a design, we only need one memory access to obtain metadata to determine the next data block and narrow down the last-mile search range to 256 bytes, effectively reducing memory accesses on real-world datasets.

\subsection{Memory Layout} \label{sec:memoryLayout}

\begin{figure}[ht]
  \vspace{-0.3cm}
  \centering
  \includegraphics[width=\linewidth]{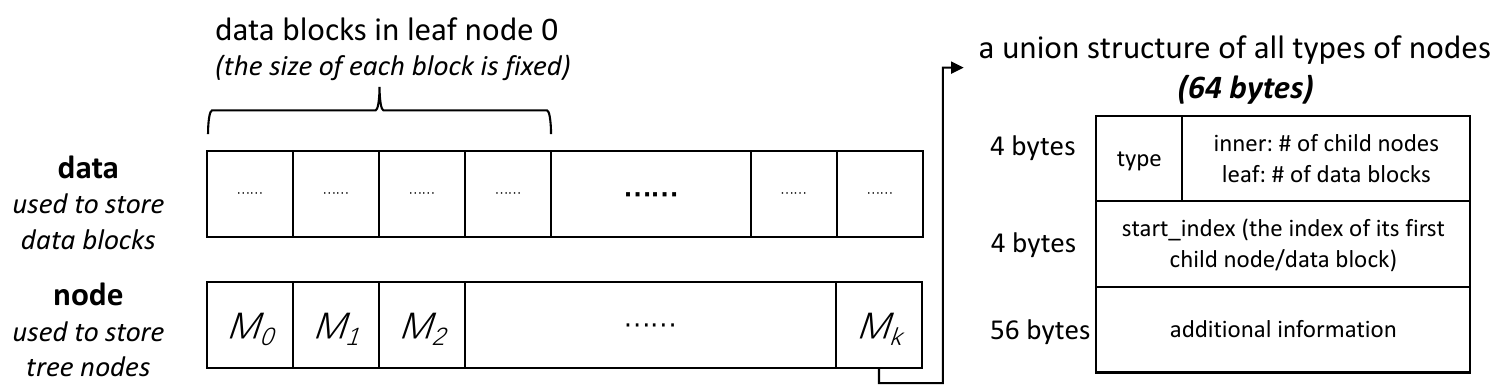}
  \vspace{-0.4cm}
  \caption{Memory Layout}
  \Description{Memory Layout}
  \vspace{-0.3cm}
  \label{fig:memoryLayout}
\end{figure}

For the memory layout of CARMI, we have two main arrays, $data$ and $node$, to assist in implementing our cache-aware design. These two arrays are used to store data points and tree nodes, respectively, as shown in Figure~\ref{fig:memoryLayout}, with details as follows:
\begin{itemize}[leftmargin=*]
  \item \textbf{Data array:} The $data$ array is used to store data points in CARMI. It is a large array containing many small data blocks, each of which has a fixed size, represented by a parameter $kBlockSize$ with a default value of 256. Data points are stored in these data blocks. In the $data$ array, data blocks managed by the same leaf node are stored in adjacent locations.
  \item \textbf{Node array:} All tree nodes, including both inner and leaf nodes, are stored in the $node$ array. Each tree node occupies a total of 64 bytes: the first byte is always the node type identifier, and the next three bytes are used to store the number of child nodes (the number of data blocks for leaf nodes). For inner nodes, the following 4 bytes represent the starting index of the child nodes in the $node$ array. For leaf nodes, they store the starting index of data blocks in the $data$ array instead. The remaining 56 bytes store additional information depending on the tree node type.
\end{itemize}

With the help of the above two arrays, we can perform operations such as lookup and insert. The way we use them in lookup and insert operations is described below.

\subsubsection{Lookup Operation} 

When accessing a data point, we first use the root node model to compute the index of the node in the next layer. Next, we access the tree node according to the index value and use its model to update the value of the index variable. This process is repeated iteratively until a leaf node is visited. Finally, we search within this leaf node to get the corresponding data record. \techreport{The pseudocode of the data access process can be found in Algorithm~\ref{alg:find}.}

\techreport{
\begin{algorithm}[ht]
	\caption{Lookup}
	\label{alg:find}
	\begin{algorithmic}[1]
		\REQUIRE ~~ $key$
    \ENSURE ~~ $data$
    \STATE $idx \leftarrow root.Predict(key)$
    \WHILE {\TRUE}
      \IF{$node[idx].type$ == $inner\_node$}
        \STATE $idx \leftarrow node[idx].Predict(key)$
      \ELSE
        \STATE /* access leaf node */
        \STATE $l \leftarrow node[idx].start\_index$
        \STATE $p \leftarrow node[idx].Predict(key)$
        \STATE $res \leftarrow SearchBlock(data[l+p], key)$
        \RETURN $data[l+p].slots[res]$
      \ENDIF
    \ENDWHILE
\end{algorithmic}
\end{algorithm}
}

\subsubsection{Insert Operation}\label{sec:insert}
\begin{figure}[ht]
  \centering
  \vspace{-0.3cm}
  \includegraphics[width=\linewidth]{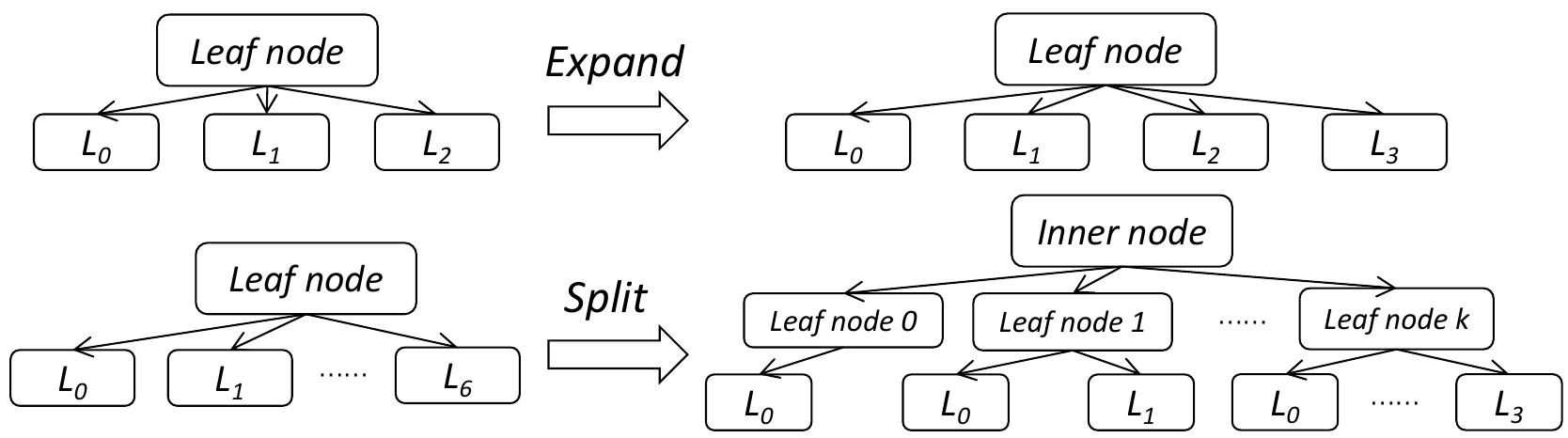}
  \vspace{-0.4cm}
  \caption{Expand and Split Mechanism}
  \Description{Expand and Split Mechanism}
  \vspace{-0.4cm}
  \label{fig:insertMechanism}
\end{figure}

The basic process of the insert operation is similar to the lookup operation. After finding the correct data block for insertion, we insert the data point into it. In addition, there are two mechanisms that can be initiated by the leaf node under certain situations:

\begin{itemize}[leftmargin=*]
  \item \textbf{Expand:} When a leaf node needs more space, it can initiate an expand operation to get more data blocks. We first collect all the data points stored in it, and then construct a new leaf node at a new location with more space, as shown in Figure~\ref{fig:insertMechanism}. The new leaf node then replaces the original one to complete the process.
  \item \textbf{Split:} If we can no longer use a single leaf node to efficiently manage all the data points, it needs to be split. The leaf node will be replaced with a subtree consisting of a new inner node and several new leaf nodes, as shown in Figure~\ref{fig:insertMechanism}. The number of leaf nodes in the subtree depends on the trained model of the new inner node.
\end{itemize}  

\subsection{Prefetch}\label{sec:prefetch}
\subsubsection{Prefetch Mechanism}
In some cases, when the key value distribution in the dataset is very regular (e.g., uniform distribution), the index of the data block can be directly predicted from the input key value. In such cases, we can further reduce the data access latency by utilizing memory prefetching. More specifically, we add an additional prefetch model to the root node to predict the data block where a given key value may be stored. Therefore, during the access process, the predicted data block is prefetched at the root node, which can be executed in parallel with other memory accesses in the normal process. If the prediction is correct, then the data block will be available in the cache when we need to access it. In this way, for datasets with regular data distribution, we can speed up the access to a certain extent. We demonstrate the prefetch mechanism via the following example:

\begin{figure}[ht]
  \centering
  \vspace{-0.3cm}
  \includegraphics[width=\linewidth]{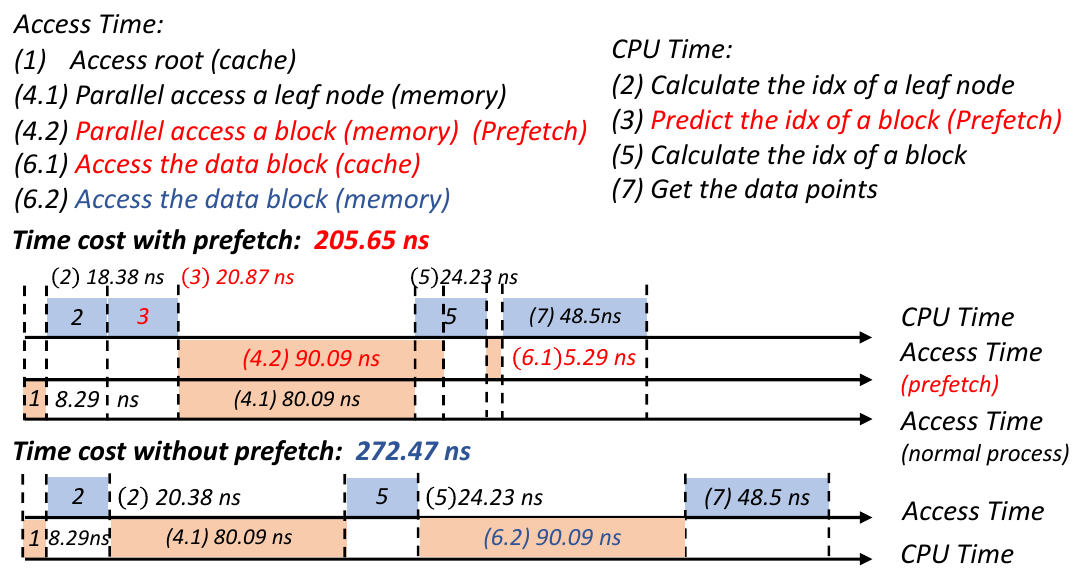}
  \vspace{-0.3cm}
  \caption{Memory Prefetch in CARMI}
  \Description{Memory Prefetch in CARMI}
  \label{fig:prefetch}
  \vspace{-0.3cm}
\end{figure}
\begin{example}
We use the same settings as in Example~\ref{example1} to show the effect of prefetching. The time cost of accessing a data point in CARMI can be divided into two parts: CPU time and access time. In this example, we need to access: the root node (8.29 ns), the leaf node (80.09 ns), and the data block (90.09ns). The CPU time consists of model calculations in the root node (20.38 ns), the leaf node (24.23 ns), and the search process (48.5 ns) in the data block. Note that the CPU calculations must wait until the corresponding cache/memory access is completed before it can be processed. Therefore, without prefetching, CPU calculations and access operations are performed alternately and do not overlap, which constitutes a total time cost of 272.47 ns.

To utilize the prefetch mechanism, we need to add an additional 28.87 ns of CPU time to predict the index of the data block that needs to be prefetched, so that we can then prefetch this data block in advance at the root node. The benefit of this prefetching step is that, when we actually need the data block later on in the process, it will be already in the cache and be retrieved very quickly, reducing the time cost of a memory access. In general, a lookup operation with prefetching only requires 205.65 ns in this example, which reduces the time by 67 ns compared to the situation without prefetching.

\end{example}

\subsubsection{Prefetch Support}
In order to support the prefetching design, we need to make a few changes to the construction algorithm: we need to add an additional model at the root node to predict the index of the data block, and we also want to store data points in data blocks according to the prefetch prediction model whenever possible. \paper{Details of these changes are not included in this paper due to page limitations and can be found in our technical report~\cite{zhang2021carmi}, together with the pseudocode of the construction algorithm.}

\techreport{
This prefetch prediction model is used to predict the index of the data block corresponding to each key value during the lookup procedure, and we store data points in data blocks according to the model during the construction of the index structure whenever possible.

\begin{figure}[ht]
	\centering
	\vspace{-0.2cm}
	\includegraphics[width=0.9\linewidth]{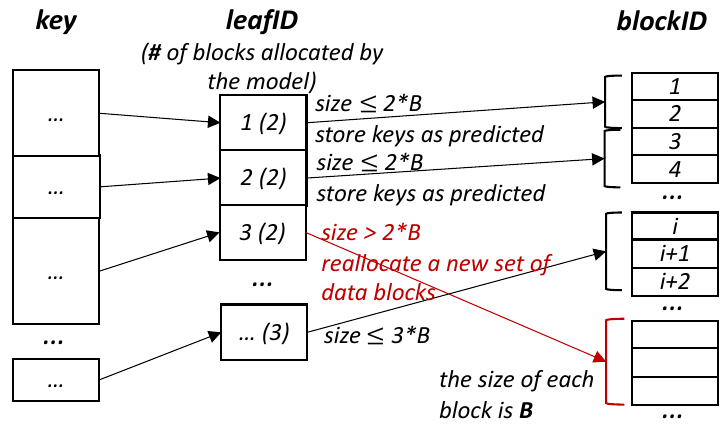}
	\caption{Prefetch Support}
	\Description{Prefetch Support}
	\label{fig:prefetchModel}
	\vspace{-0.2cm}
\end{figure}

More specifically, we use a piecewise linear regression model as the prefetch prediction model. The prefetch prediction model requires access to the raw output of the root model (leaf index before rounding down), and then use it as input to the piecewise linear model to compute a block index. Note that by the time we start to construct the prefetch prediction model, the root node has already been determined by the greedy node selection algorithm. Therefore, the unrounded leaf node index can be directly obtained from the root node.

In the piecewise linear regression model, we force the slope and intercept of each segment to be integers, so that within each segment, each leaf node is mapped to the same number of data blocks. For each leaf node, we first attempt to allocate its data blocks according to the prefetch prediction model, and check whether the current leaf node has enough capacity to store all assigned data points. If the leaf node has enough capacity, then we simply store the data points in these data blocks. Otherwise, we reallocate a new set of data blocks for the current leaf node at another irrelevant location. Figure~\ref{fig:prefetchModel} illustrates the process of data block allocation for supporting the prefetch mechanism.

Within each leaf node, we try to store as many data points in the predicted data block as possible, to increase the success rate of prefetch. This corresponds to another optimization problem within the leaf node.

\subsubsection{The Prefetch Prediction Model}
To learn the prefetch prediction model, we also establish a cost model for it and optimize accordingly. Let $L = \{M_1, \ldots,$ $M_{n}\}$ be the collection of leaf nodes directly under the root node, then the cost of index accesses on these nodes can be approximated as:
\begin{align} \label{eq:prefetch}
	\sum_{i=1}^{n} [\frac{\left| Q_{M_i} \right|}{\left| Q \right|}\mathbbm{E}_{q \in Q_{M_i}} [TCost(M_i, q)] + \lambda SCost(M_i)]
\end{align}
where $Q_{M_i}$ is the collection of queries accessing $M_i$.

\begin{figure*}[ht]
	\vspace{-0.8cm}
  \centering
  \setlength{\abovecaptionskip}{-3pt}
  \includegraphics[width=0.95\linewidth]{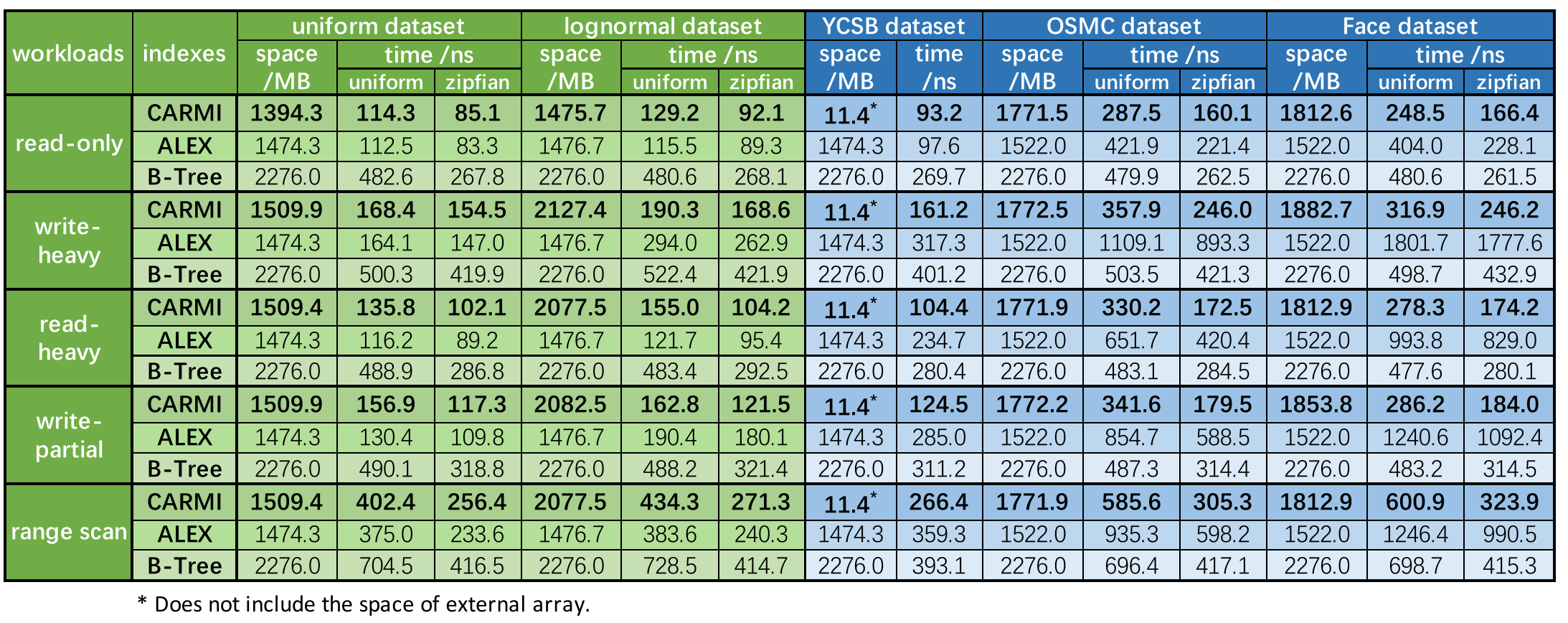}
  \caption{CARMI vs. Baselines: Time and Space Usage Comparison.}
  \Description{CARMI vs. Baselines: Time and Space Usage Comparison.}
  \label{fig:totalRes}
  \vspace{-0.3cm}
 \end{figure*}
}

Considering the effect of prefetching, the time cost of data blocks becomes slightly different.  a leaf node that supports prefetch (i.e., allocated to its predicted location), the time cost of accessing data blocks is now equal to the latency of cache memory, and the value of $TCost(M_i, q)$ also changes accordingly. This downside is that now $SCost(M_i)$ can be slightly larger than needed.

We use a dynamic programming algorithm to optimize Equation~\ref{eq:prefetch}. The DP state $cost[i,r]$ is the total cost of first $r$ leaf nodes when we use $i$ segments to model them. This is a standard 2D dynamic programming and we omit the algorithm details here.

\section{Experiments}\label{sec:experiment}
In this section, we conduct experiments on various datasets and workloads to evaluate the performance of CARMI and delve into CARMI from different aspects through several auxiliary experiments. \paper{Due to page limitations, some contents, such as statistics of tree structures, are described in our technical report~\cite{zhang2021carmi}.}

\paper{
  \begin{figure*}[ht]
  \centering
  \vspace{-0.6cm}
  \includegraphics[width=0.98\linewidth]{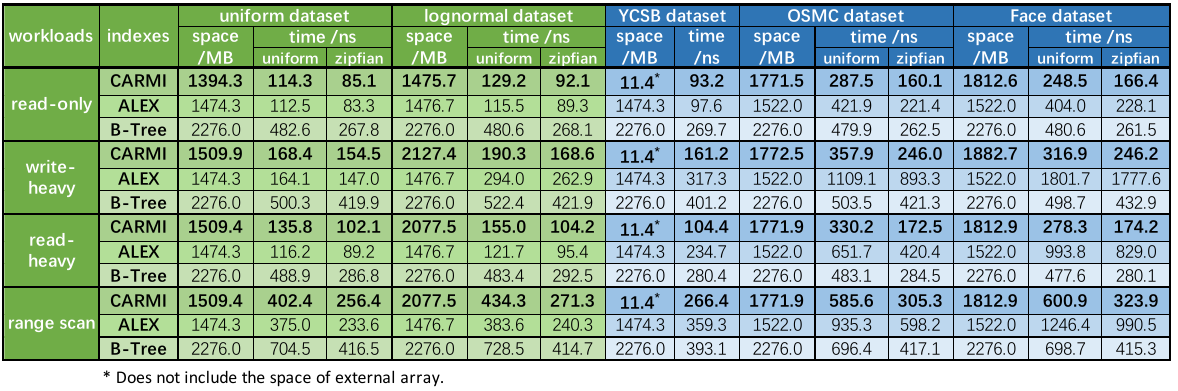}
  \vspace{-0.3cm}
  \caption{CARMI vs. Baselines: Time and Space Usage Comparison.}
  \Description{CARMI vs. Baselines: Time and Space Usage Comparison.}
  \label{fig:totalResPaper}
  \vspace{-0.3cm}
\end{figure*}
}

\subsection{Experimental Setup}\label{sec:setup}
We conducted all the single-threaded experiments on an Ubuntu Linux machine equipped with an AMD Ryzen 3700X 8-Core Processor and 32GB RAM.

\subsubsection{Datasets}
Seven datasets are used in our main experiments. Their details are listed as follows:
\begin{itemize}[leftmargin=5mm]
  \item \textbf{synthetic dataset:} $4$ synthetic datasets are generated from $4$ different distributions: (a) \textbf{lognormal}: $log(key) \sim \mathcal{N}(0, 1)$; (b) \textbf{uniform}: $key \sim \mathcal{U}(0, 1)$; (c) \textbf{normal}: $key \sim \mathcal{N}(0, 1)$; (d) \textbf{exponential}: $key \sim \text{Exp}(0.25)$. All these datasets are stored as key-value pairs of <$double, double$>. The size of each dataset is 1GB, and the key values of data points are multipled by $10^8$.
  \item \textbf{YCSB dataset}: This dataset consists of 67,108,864 data records generated by YCSB benchmark~\cite{cooper2010benchmarking}. We build an index on the $YCSB\_KEY$ attribute, which obeys a uniform distribution.
  \item \textbf{OSMC dataset:} This dataset is generated from a public dataset of Open Street Maps~\cite{OpenStreetMap}, which includes the latitude and longitude of points around the world. The number of data points in this dataset is 67,108,864 (the size of each dataset is 1GB). This dataset is stored as key-value pairs of <$uint64, uint64$>.
  \item \textbf{Facebook dataset:} This dataset is generated from the IDs of Facebook users~\cite{van2019efficiently, marcus2020benchmarking}, consisting of 67,108,864 data points. This dataset is stored as key-value pairs of <$uint64, uint64$>.
\end{itemize}

We have also ported CARMI into the SOSD\techreport{~\cite{marcus2020benchmarking}}\paper{~\cite{sosd-vldb}} platform to test its performance and the details are shown in Section~\ref{sec:sosd}.

 \subsubsection{Evaluation Workloads}\label{sec:workload}
 We use \paper{4}\techreport{5} different query workloads in our experiments. The first three workloads are similar to the workloads in the YCSB benchmark~\cite{cooper2010benchmarking}: (a) a write-heavy workload with a mix of 50\% reads and 50\% inserts; (b) a read-heavy workload with a mix of 95\% reads and 5\% inserts; (c) a read-only workload. \techreport{In addition, we include a write-partial workload in which insert operations are concentrated in a small key value range. This workload consists of 85\% reads and 15\% inserts, and the inserted data points are concentrated between 60\% and 90\% of the sorted dataset. The write-partial workload is intended to mimic some real-world scenarios in which the keys of new data points are contained in a small value range during each session (e.g., when using the current date/time as key value).} Finally, we include a range scan workload with a mix of 95\% range scan and 5\% inserts as in~\cite{ding2019alex}.

 For each workload, we exectue 100,000 operations and measure the average time used by each operation. For all workloads, lookup keys are selected at random from the existing keys in the index. We consider two access patterns for generating queries: a Zipfian distribution (the normalized frequency of the $x$-th element is: $f(x) = \frac{1}{x^\alpha} / \sum_{i=1}^N \frac{1}{i^\alpha}$, $\alpha = 0.99$ as in~\cite{ding2019alex}) and a uniform distribution. In workloads involving insert operations, read and insert operations are performed alternately in proportion. For example, in read-heavy workloads, we execute 19 lookups followed by 1 insert operation. For range scan workloads, the length of each range scan is uniformly sampled from [1, 100].

 For the YCSB dataset, the queries are handled differently to simulate real-world use cases: read queries are generated from a Zipfian distribution as usual, but insert keys are monotonically increasing to be consistent with the YCSB benchmark.

\subsubsection{Implementation and Baseline}\label{sec:paramsSetting}

We compare against the following baselines: STX B+ Tree~\cite{STXBtree} and ALEX~\cite{ding2019alex}. The node size of B+ Tree is 512 bytes, which is optimal for in-memory queries~\cite{zhang2016reducing}. All other parameters use the default values in the source code. SOSD benchmark~\cite{marcus2020benchmarking} also includes some other indexes as baselines in Section~\ref{sec:sosd}. CARMI is open-sourced and can be found on Github~\cite{carmiGithub}. In our experiments, we only tune one parameter $\lambda$, which is sensitive to data distribution and dataset size, and all other parameters do not need to be tuned and use the default values unless otherwise stated. \paper{Among them, the size of each data block is 256 bytes, and the maximum capacity of external leaf nodes is 512. The default value of $kDPThreshold$ is 512, which is used to switch between the DP and greedy algorithms. If the size of a sub-dataset is less than 90, the algorithm will directly construct a leaf node instead of choosing a better one from leaf/inner nodes. When a leaf node needs to split, we replace it with an inner node and 16 leaf nodes.} \techreport{More details can be found in the appendix.}

\subsubsection{Training Queries for Index Construction in CARMI}

In our experiments, the training query workload for index construction is specified as a uniform read access over all data points, and a uniformly sampled subset of data points\techreport{\footnote{The insert queries in the training queries of the write-partial workload are sampled between 60\% and 90\% of the overall dataset.}} to serve as key values for insert queries (with the exception of YCSB dataset). The ratio of read/insert queries is the same as in the target query workload.

\subsection{General Efficiency Comparison}\label{sec:general}
  In this section, we evaluate the performance of CARMI and compare it with the baselines. We consider 5 query workloads and 7 different datasets, as explained in Section~\ref{sec:setup}. We consider two access patterns for datasets other than YCSB: uniform and Zipfian, as illustrated in Section~\ref{sec:workload}. These result in a total of 65 possible configurations. \techreport{Figure~\ref{fig:totalRes}} \paper{Figure~\ref{fig:totalResPaper}} shows the results for 45 of them due to space limitation, and the results for normal and exponential datasets are omitted since they are similar to those of the uniform dataset. 
  
  In general, the time efficiency of both learned indexes is significantly better than B+ Tree. Comparing CARMI with ALEX, which is also a learned index, we see that these two learned indexes perform roughly the same over synthetic datasets. However, CARMI significantly outperforms on these real-world datasets, achieving an average speedup of 1.2$\times$/2.2$\times$ on read-only/read-write workloads. This demonstrates the effectiveness of data partitioning and cache-aware design over real-world datasets, in which data location is much harder to predict.

\subsubsection{Read-only Workload}\label{sec:readonlyResult}
  For read-only workloads, the lookup speed of CARMI is about 1.6-4.2$\times$ faster than B+ Tree, and 1.2$\times$ faster on average than ALEX. Meanwhile, CARMI uses only 0.7$\times$ of memory space compared to B+ Tree (excluding YCSB\footnote{For the YCSB dataset, since CARMI uses the external leaf nodes, which only store a pointer of the location of data points, the space cost of 11.4 MB only counts the space of the tree structure itself and does not include the space used by external data.} dataset).

  It is of interest to compare CARMI with B+ Tree, since their difference is only on the upper level. The significant speed gain of CARMI is mainly due to the following reasons: (a) CARMI uses fast model prediction instead of binary search. (b) the larger fanout of the root nodes in CARMI reduces the depth of the index. Then, most of the data points are managed by a leaf node that is directly under the root node, making the index structure flatter. To verify that our construction algorithm can build a flatter index, we also calculated the average tree depth with respect to keys, where a single root node has a depth of 1. The average tree depth of CARMI is 2.1, while that of ALEX and B+ Tree are 2.5 and 7.3, respectively.

\begin{figure*}[ht]
    \centering
    \vspace{-0.8cm}
    \includegraphics[width=\linewidth]{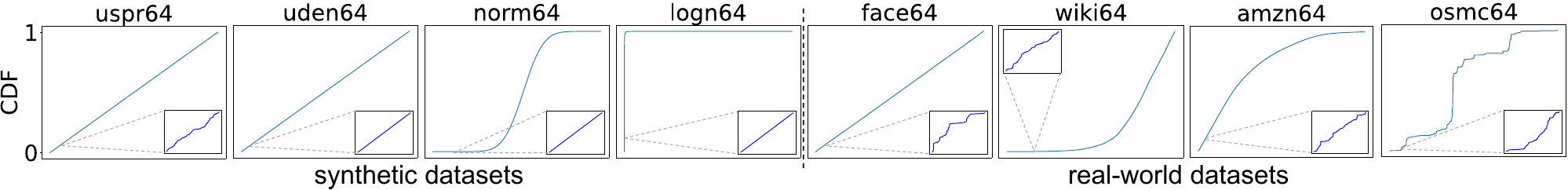}
    \vspace{-0.6cm}
    \caption{The CDFs of Datasets in SOSD~\cite{marcus2020benchmarking}.}
    \Description{The CDFs of Datasets in SOSD~\cite{marcus2020benchmarking}.}
    \label{fig:cdf}
    \vspace{-0.3cm}
  \end{figure*}

  Comparing the two learned indexes, the main difference is in the performance over OSMC/Face, in which CARMI achieves an average speedup of 1.5$\times$. The two datasets are real-world datasets with highly non-linear local distribution (the CDFs can be found in Figure~\ref{fig:cdf}), which invalidates ALEX's strategy of storing data points according to the predicted location, and causes the index to grow deeper. Meanwhile, our hybrid construction algorithm can automatically select suitable nodes to obtain better performance. Section~\ref{sec:sosd} shows similar results on other real-world datasets as well, making it clear that this is not a coincidence.

  In summary, CARMI outperforms B+ Tree and can achieve similar or better performance than ALEX under read-only workloads, and the gain is more evident on real-world datasets.

\techreport{
 \subsubsection{Write-heavy Workload}
  CARMI achieves an average speedup ratio of 2.2$\times$ compared to B+ Tree and 3.0$\times$ compared to ALEX\footnote{The insert of the YCSB protocol is to continuously insert new data points at the end of the dataset, and ALEX does not strictly follow this mode for insert in their paper. Thus, our reported number is different compared to the values in their paper.} on write-heavy workloads.
  
  It is generally difficult for learned indexes to divide non-linear datasets evenly, leading to a large gap between the numbers of data points in each node. This results in additional expansion costs in ALEX during the insert operations. However, CARMI handles well even under such scenarios, using only 0.42$\times$ time of ALEX on lognormal/real-world datasets. As for space cost, CARMI uses 1.2$\times$ space compared to ALEX, and 0.8$\times$ compared to B+ Tree on average.

  In summary, CARMI has a better time efficiency and uses a similar amount of space compared to ALEX, and the gap is more significant on non-linear and real-world datasets.

 \subsubsection{Other Read-Write Workloads}
  CARMI has a shorter access time (2.04$\times$/2.03$\times$ average speedup comapred to B+ Tree and ALEX respectively) and use less space (0.79$\times$ on average compared to B+ Tree) for all other read-write workloads including range scan workload. This shows that CARMI can better handle various workloads, especially non-linear datasets.
}

\subsection{SOSD Results}\label{sec:sosd} 
 In order to further test the performance of CARMI on real-world datasets, we have integrated CARMI into the SOSD benchmark~\cite{marcus2020benchmarking,sosd-vldb}, a platform for testing learned index structures. We use all real-world datasets in SOSD, each of which consists of 200 million unsigned 32-bit/64-bit integers: \textbf{amzn} is book sale popularity data~\cite{Amazon}, \textbf{face} is the IDs of Facebook users~\cite{van2019efficiently}, \textbf{wiki} is Wikipedia article edit timestamps~\cite{wiki}, and \textbf{osmc} is generated from a public dataset of Open Street Maps~\cite{OpenStreetMap}. SOSD performs 10 million lookups on each dataset, where the lookup keys are uniformly chosen from the set of keys, and computes the average latency per lookup.

 There are 10 baselines in SOSD and are categorized into four types: (a) \textbf{three learned indexes}: RMI, RS and ALEX; (b) \textbf{three traditional indexes}: B+ Tree, FAST and ART; (c) \textbf{three on-the-fly} algorithms that directly operate on a sorted array: BS, IS and TIP; (d) \textbf{one auxiliary index} that uses small auxiliary structures: RBS. Among them, FAST and ART do not support all the datasets as explained in SOSD and more details can be found in~\cite{marcus2020benchmarking}. \paper{Table~\ref{tab:sosdPaper} shows the average lookup time of indexes on \textbf{all real-world datasets} in the SOSD benchmark, where the results with the shortest latency and those slightly slower (within 20ns) are bolded.} \techreport{Table~\ref{tab:sosdPaper} shows the average lookup time of indexes on \textbf{all real-world datasets} in the SOSD benchmark, where the results with the shortest latency and those slightly slower (within 20ns) are bolded.} 
 
  \begin{table}[ht]
   \center
   \small
   \caption{The Average Lookup Time (ns) on SOSD Platform.}
   \vspace{-0.3cm}
   \label{tab:sosdPaper}
   \begin{tabular}{c|cc|cccc}
     \toprule 
     ~&\multicolumn{2}{c|}{uint32}&\multicolumn{4}{c}{uint64}\\
     (ns)&amzn &face &amzn&face&wiki&osmc\\
     \midrule
     CARMI&\textbf{192.84} &\textbf{187.43} &\textbf{201.80} & \textbf{334.41}& \textbf{217.50}&\textbf{368.65} \\
     \midrule
     RMI& 265.43&274.53 &266.54 &\textbf{334.54}  &\textbf{222.43} &402.32  \\
     RS& 280.03 & 362.54 &296.61& 436.63 &\textbf{218.43}&412.39 \\
     ALEX&\textbf{210.99} &434.69  &251.32 & 496.48&  289.81&499.04 \\
     \midrule
     RBS& 325.43 &312.39  &385.96 &\textbf{334.73} &335.75 &529.06   \\
     \midrule
     FAST& 246.03&228.79   &N/A&N/A&N/A &N/A  \\
    ART&N/A&\textbf{182.36} &N/A&391.76&N/A&N/A  \\
     B+Tree& 529.43&524.54 &601.23&592.43 &608.42 &599.43 \\
     \midrule
     BS& 1014.5&983.49 &1015.1&961.93 &1002.3&987.24 \\
     TIP& 731.97&880.12 &750.89&1124.6 &942.84 &4773.8 \\
     IS& 3852.7&1007.7 &4103.9&1494.8 &6836.4&66474\\
     \bottomrule
   \end{tabular}
 \end{table}
  

We also examine the distribution of datasets. As shown in Figure~\ref{fig:cdf}, a major difference between synthetic and real-world datasets is that synthetic datasets all have locally linear CDF, which is rare in real-world datasets. The highly non-linear local CDFs of real-world datasets make it difficult for prior solutions to predict the location of individual data records accurately, leading to large-range searches in the last mile. In contrast, because CARMI is designed based on the data partitioning view with a more cache-friendly leaf node layout, it is less penalized by these highly non-linear parts and thus performs better in real-world datasets: CARMI can achieve an average speedup of 1.21$\times$ even compared to a well-tuned RMI.

Interestingly, most learned indexes do not perform better than traditional indexes on particular real-world datasets, as shown in Table~\ref{tab:sosdPaper}. Specifically, FAST takes only 237.41 ns on average on amzn32 and face32, while RMI, RS, and ALEX take 1.14$\times$/1.35$\times$/1.36$\times$ average lookup time, respectively. ART takes an average of 287.06 ns on face32 and face64, while RMI, RS, and ALEX take 1.06$\times$/1.39$\times$/1.62$\times$, respectively. Despite that, the lookup latency of CARMI is robust enough to be the shortest or very close to the shortest among all indexes for all datasets.



We also report the average space usage of each index in Table~\ref{tab:sosdspace}. The space required by CARMI is comparable to that of traditional indexes with good performance, such as FAST and ART. Note that the space cost of CARMI can be adjusted by users.

  \begin{table}[ht]
   \center
   \small
   \caption{Average Space Usage (MB) on SOSD.}
   \vspace{-5pt}
   \label{tab:sosdspace}
   \begin{tabular}{c|cccccc}
     \toprule 
     ~&CARMI &RMI&RS&ALEX&\multicolumn{1}{|c}{RBS}\\
     \midrule
     uint32& 3918.30 & 2471.65&2299.63 &3336.74 &\multicolumn{1}{|c}{2289.82} \\
     uint64& 5481.44&3143.31&3064.33&4430.83&\multicolumn{1}{|c}{3052.76}\\
     \midrule
    ~& FAST&ART&B+Tree&\multicolumn{1}{|c}{BS}&TIP&IS\\
     \midrule
     uint32& 5120.00 &4596.52 &2657.31 &\multicolumn{1}{|c}{2288.82} & 2288.82&2288.82\\ 
     uint64& N/A &5280.92  &3540.52 &\multicolumn{1}{|c}{3051.76} & 3051.76 &3051.76\\
     \bottomrule
   \end{tabular}
  \vspace{-0.1cm}
  \end{table}

\subsection{Tradeoff between Time and Space}\label{sec:tradeoff}
\begin{figure}[ht]
  \centering
  \vspace{-0.3cm}
  \includegraphics[width=\linewidth]{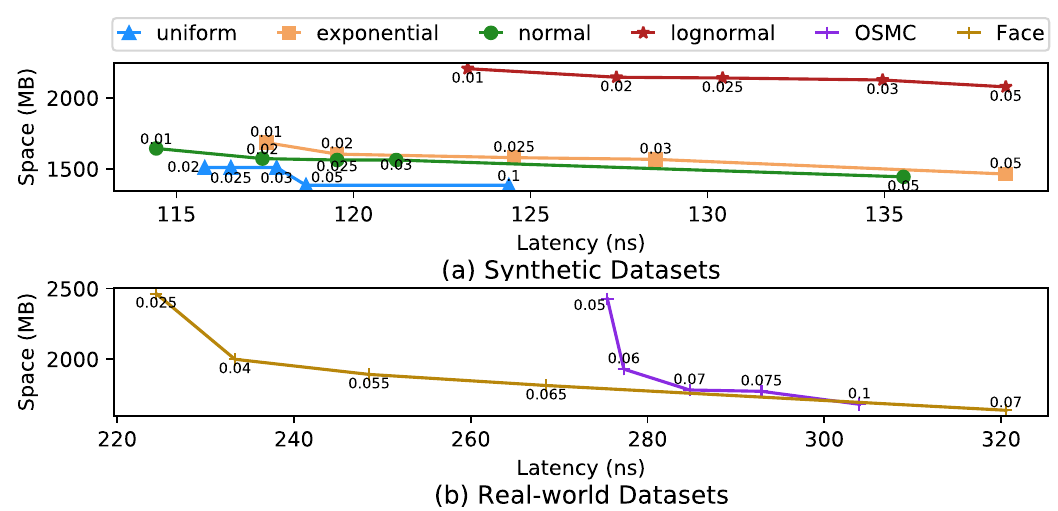}
  \vspace{-0.3cm}
  \caption{The Tradeoff Between Time and Space Cost}
  \Description{The tradeoff between time and space}
  \label{fig:lambda}
  \vspace{-0.4cm}
\end{figure}

One additional advantage of our cost-based index construction algorithm is that one can now flexibly choose the balance point between time and space cost. As shown in Figure~\ref{fig:lambda}, by tuning the value of parameter $\lambda$ in Problem~\ref{prob:linear}, one can reduce the memory usage of indexes at the cost of increased read access latency. The curves of real-world datasets exhibit the characteristics of convex functions, but the optimal solution for each dataset depends on the actual data distributions. Therefore, we need to carefully tune $\lambda$ according to practical scenarios to find the optimal solution, so as to obtain a more desired tradeoff between time and space.

\techreport{  
  \begin{table*}[ht]
    \center
    \small
    \vspace{-0.3cm}
    \caption{Statistics on Different Root Fanouts.}
    \label{tab:ablationMemory}
    \vspace{-0.3cm}
    \begin{tabular}{c|ccc|ccc|ccc|ccc}
      \toprule  
\multirow{2}*{Root}&\multicolumn{3}{c|}{Uniform Dataset}&\multicolumn{3}{c|}{Normal Dataset}&\multicolumn{3}{c|}{Lognormal Dataset}&\multicolumn{3}{c}{OSMC Dataset}\\
~&\multirow{1}*{Time}&\# of&\multirow{2}*{Percentage}&\multirow{1}*{Time}&\# of&\multirow{2}*{Percentage}&\multirow{1}*{Time}&\# of&\multirow{2}*{Percentage}&\multirow{1}*{Time}&\# of&\multirow{2}*{Percentage} \\
Fanout&/ns&Accesses&~&/ns&Accesses&~&/ns&Accesses&~&/ns&Accesses&~ \\
      \midrule
      65536  &301.64&6.82&5.14\%&355.39&7.67&4.76\%&372.72&8.73&3.32\%&593.10&11.95&0.22\%\\
      131072 &280.58&5.72&7.28\%&333.46&6.66&6.74\%&337.73&7.69&4.68\%&539.83&11.15&0.38\%\\
      262144 &254.16&4.62&10.36\%&300.26&5.65&9.58\%&308.69&6.64&6.62\%&487.83&10.33&0.67\%\\
      524288 &212.34&3.52&14.66\%&276.09&4.65&13.55\%&283.86&5.58&9.35\%&449.08&9.52&1.14\%\\
      1048576&168.43&2.47&20.66\%&232.51&3.68&19.05\%&261.56&4.54&13.05\%&410.11&8.69&1.89\%\\
      2097152&152.71&1.53&28.64\%&212.36&2.80&26.25\%&230.68&3.53&17.99\%&374.43&7.85&3.05\%\\
      \bottomrule
    \end{tabular}
  \end{table*}
}

\subsection{Cost of Construction}\label{sec:cost}
\begin{figure}[ht]
    \centering
    \vspace{-0.3cm}
    \includegraphics[width=8cm]{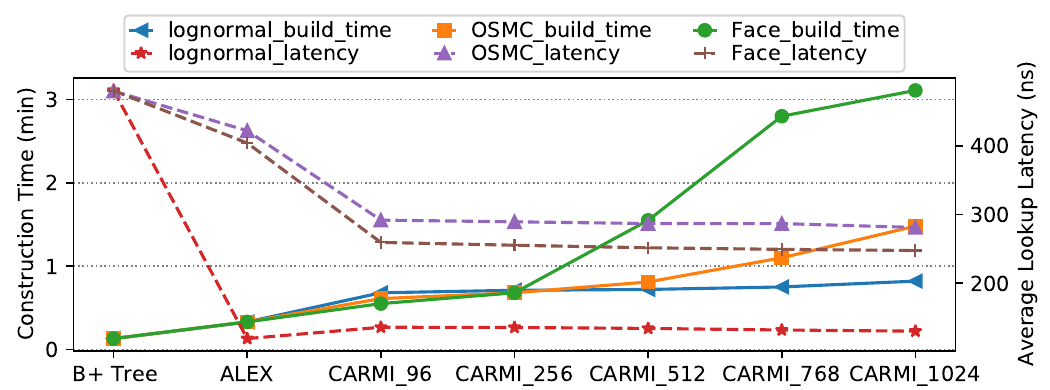}
    \vspace{-0.3cm}
    \caption{Construction/Lookup Time of Indexes.}
    \Description{Construction/Lookup Time of Indexes.}
    \label{fig:time}
    \vspace{-0.3cm}
  \end{figure}

\paper{\revision{We compare the time to construct each index for 1GB datasets using different values of the parameter $kDPThreshold$. This parameter is used to adjust the threshold for the size of the sub-datasets using the DP algorithm, thus obtaining a tradeoff between the construction time and the average lookup latency. As shown in Figure~\ref{fig:time}, the construction of CARMI with different values of $kDPThreshold$ can be finished within 0.3-3.2 minutes, while B+ Tree and ALEX take 10s and 20s on average, respectively. Although CARMI takes longer to build indexes than baselines, it is generally acceptable for most practical scenarios. Moreover, $kDPThreshold$ has a slight impact on the average lookup latency, with the difference between 96 and 1024 being around 10 ns.}
}

\techreport{
  The construction of the index structure should only be performed periodically when the database system has enough computation resources to spare (e.g., during night time when the frequency of the incoming queries is low). As such, the time efficiency of the construction algorithm is not a major concern. Nevertheless, we still want to have a basic understanding of its time cost.

  Recall that we have two algorithms (the greedy algorithm and the DP algorithm) to construct a node. In order to achieve a balance between the construction time and the average lookup latency, we define a parameter $kDPThreshold$ in CARMI, and use the greedy algorithm when the size of the sub-dataset is greater than the parameter. As shown in Figure~\ref{fig:time}, the construction of CARMI with different values of $kDPThreshold$ can be finished within 0.3-3.2 minutes, while B+ Tree and ALEX take 10s and 20s on average, respectively. Although CARMI takes longer to build indexes than baselines, it is generally acceptable for most practical scenarios. Moreover, the value of this parameter has a slight impact on the average lookup latency, with the difference between a value of 96 and 1024 being around 10 ns.
  
  To conclude, in scenarios where construction time is more important, $kDPThreshold$ can be tuned to reduce the construction time. While in most practical scenarios, the default value can be used directly.
}

  \paper{\revision{To conclude, in scenarios where construction time is important, $kDPThreshold$ can be tuned to reduce the construction time. While in most practical scenarios, the default value can be used directly.}
  }

\techreport{
  \subsection{The Effect of Fanout}\label{sec:fanout}
  In order to show the effect of root fanout on average access time, we use CARMI to simulate a two-layer static RMI with adjustable root fanout. The root node, which is the linear regression model, directly manages the leaf nodes. Each leaf node is an external array leaf node. Table~\ref{tab:ablationMemory} shows the average access time and the average number of memory accesses required per query under different root fanouts. We also count the percentage of data points that are stored in the same location as the location calculated by the linear model across all data points.

  As demonstrated in Table~\ref{tab:ablationMemory}, the average access time tends to be smaller as the fanout increases. The gains are particularly evident on non-linear datasets like the OSMC dataset. The benefits brought by a large fanout are mainly due to the following reasons: (a) It reduces the number of data points in each leaf node and thereby avoids the occurrence of a wide range of binary searches, resulting in the reduced number of memory accesses. (b) It can simplify the situation of lower-level nodes, so that more data points can have their actual storage positions to be consistent with the calculation results of the linear model.

  In general, a larger fanout of the root node can reduce the average access time to some extent. Its contribution to the local efficiency of the root node is quantified in our entropy metric, which further validates our view of data partitioning.
}

  \subsection{Performance Breakdown}\label{sec:breakdown} 
  \techreport{
    In this section, we examine the performance improvements brought by each design.
    \subsubsection{Detailed Performance Study}
    First, we investigate the contribution of each idea to the overall performance. We implement 4 variants of CARMI to simulate indexes with different ideas and perform read-only workloads on 5 datasets, including 3 synthetic datasets and 2 real-world datasets. A brief introduction of these variants are as follows:

  }
    \paper{In this section, we investigate the contribution of each idea to the overall performance. We implement 4 variants of CARMI to simulate indexes with different ideas and perform read-only workloads on 5 datasets, including 3 synthetic datasets and 2 real-world datasets. A brief introduction of these variants are as follows:}

      \begin{itemize}[leftmargin=2mm]
        \item \textbf{RMI}: a two-layer RMI with an LR root node and 131072 external leaf nodes.
        \item \textbf{Greedy}: a dynamic index constructed by the greedy algorithm.
        \item \textbf{Greedy cache-aware}: a dynamic index equipped with the cache-aware design and greedy algorithm.
        \item \textbf{CARMI}: a complete CARMI equipped with all proposed ideas.
      \end{itemize}

      \begin{figure}[ht]
        \centering
        \includegraphics[width=8cm]{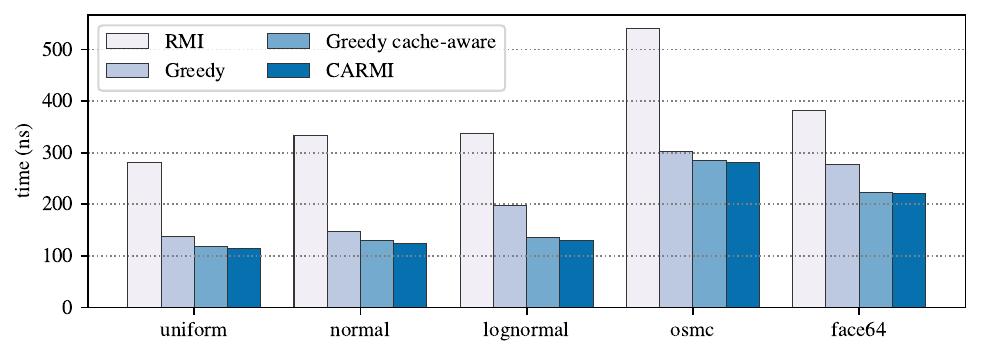}
        \vspace{-0.3cm}
        \caption{Latency of Four CARMIs.}
        \Description{Latency of Four CARMIs.}
        \label{fig:breakdown}
        \vspace{-0.3cm}
      \end{figure}

      As shown in Figure~\ref{fig:breakdown}, most of the advantages of the CARMI framework come from the data partitioning view, which enables the greedy index to achieve an average speedup of 1.84$\times$ compared to RMI. Specifically, only the greedy algorithm is used to build the index, so each node setting is determined by the entropy-based notion of local efficiency. This means that nodes can have larger fanouts for smaller sub-datasets, reducing large-range searches in RMI. In addition, four inner nodes can be mixed to handle different situations well to obtain better data partitioning effectiveness.

      Next, in the greedy cache-aware index, we replace leaf nodes with CF array leaf nodes and cooperate with the prefetching mechanism to examine the effect of our cache-aware design. Due to the influence of data distributions and different sub-datasets partitioned by the greedy algorithm, the cache-aware design shows different effects. Among them, the speedup for lognormal and face64 datasets is more significant. In summary, this index still obtains an average speedup of 1.21$\times$ compared to the greedy index, which is the contribution of cache-aware design.

      If resources are sufficient, users can use the hybrid construction algorithm consisting of the greedy algorithm and DP algorithm to build better indexes. According to the average access time when parameter $kDPThreshold$ is 1024, as shown in Figure~\ref{fig:breakdown}, the complete CARMI can reduce the average access time by about 10ns.

\techreport{
    \subsubsection{Model Characteristics}\label{sec:model character} 
      Next, we investigate the effect of using four inner nodes flexibly in the index. First, we conduct experiments on the four inner nodes to show their respective characteristics. In these experiments, the root nodes are all piecewise linear nodes, but we only allow one kind of node to be used as inner nodes. In addition, the results of CARMI are also shown in Table~\ref{tab:ablationModel}, where the hybrid algorithm flexibly uses these four inner nodes.

      \begin{table}[ht]
        \center
        \small
        \vspace{-0.2cm}
        \caption{Latency on Different Model Types.}
        \label{tab:ablationModel}
        \vspace{-0.3cm}
        \begin{tabular}{c|cccc|c}
          \toprule  
          Latency /ns&LR  &P. LR  &Hist  &BS  &CARMI\\
          \midrule
          Uniform& 136.8& 143.5& 141.3&145.8&136.6\\ 
          Normal&153.6&152.5&158.7&154.9&139.4\\ 
          Logn&157.5&155.0&155.8&153.8& 139.9\\
          OSMC&415.8 &405.3 &413.32 &374.9 & 293.4\\
          \bottomrule
        \end{tabular}
      \end{table}

      As shown in Table~\ref{tab:ablationModel}, CARMI outperforms the indexes formed by using these nodes individually, indicating that the mixed-use of these four different types of models can effectively speed up the query response. More specifically, on the one hand, LR nodes can handle evenly distributed parts well, thus beating other models on the uniform dataset. On the other hand, for the non-linear part, the PLR and Hist nodes sacrifice some model calculation time in exchange for better data partitioning results. As for the extreme case, BS node can handle it well. Although the model computation takes more time, as shown in Table~\ref{tab:dimensions}, it can partition the OSMC dataset more evenly, resulting in a shorter average access time.

      In summary, it makes sense to combine different nodes together and can be easily achieved based on our fixed-size node design. CARMI benefits from our hybrid construction algorithm and cost model to select optimal node settings in different situations to get the shortest average access time.
}

\techreport{
    \subsubsection{Prefetching Mechanism} 
      In order to evaluate the impact of the prefetching mechanism, we conduct experiments with and without prefetching instructions on these four datasets and show the time/space costs and the proportion of data blocks that are successfully prefetched in Table~\ref{tab:prefetch}. On these synthetic datasets, the average success rate of the prefetching mechanism is 93.1\%, and the average access time is reduced by 63.1 ns. For real-world datasets, most data points require at least two inner nodes to handle the data distribution, so there is little gain in prefetching them at the root node. As we discussed in Section~\ref{sec:prefetch}, the prefetching mechanism requires more space to better exert its effects.

      \begin{table}[ht]
        \center
        \small
        \vspace{-0.2cm}
        \caption{Results With/Without Prefetch Mechanism.}
        \label{tab:prefetch}
        \vspace{-0.3cm}
        \begin{tabular}{c|ccc|cc}
          \toprule  
        ~&\multicolumn{3}{c|}{with prefetch}&\multicolumn{2}{c}{without prefetch}\\
        ~&time/ns& space/MB&proportion &time/ns& space/MB\\
          \midrule
          Uniform&114.0& 1509.3&98.3\%& 185.8&1230.6\\ 
          Normal&124.6&1561.6&89.1\%&184.4&1334.2 \\ 
          Logn&126.2&2137.8&91.9\%&183.7&1912.4 \\
          OSMC&284.5&1557.86&2.5\%&289.3&1533.4\\
          \bottomrule
        \end{tabular}
      \end{table}
}

\subsection{Robustness of CARMI}\label{sec:robust}
\begin{figure}[ht]
  \vspace{-0.3cm}
  \centering
  \includegraphics[width=\linewidth]{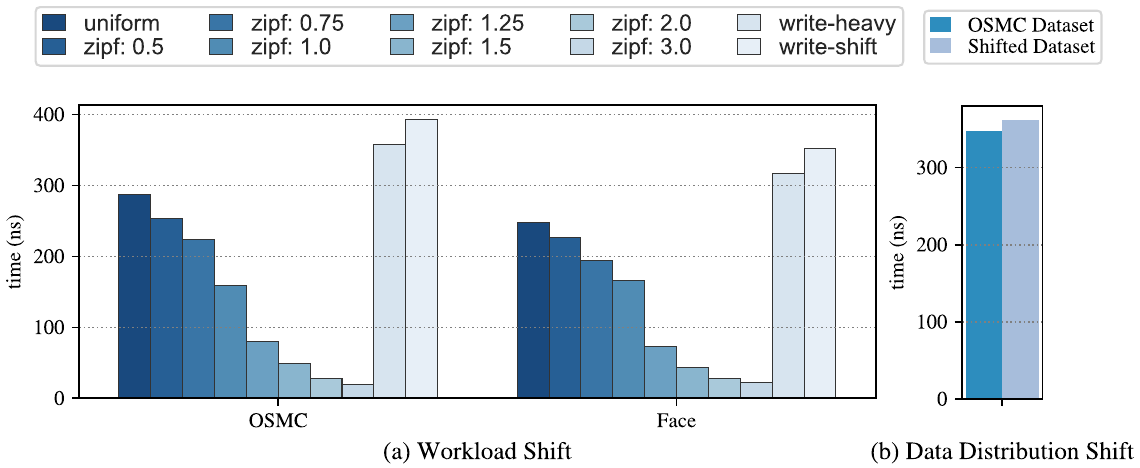}
  \vspace{-0.3cm}
  \caption{Robustness of Indexes in Different Situations.}
  \Description{Robustness of Indexes in Different Situations.}
  \label{fig:drift}
  \vspace{-0.3cm}
\end{figure}

\paper{\revision{In this section, we simulate three groups of workload and data distribution shifts to examine the robustness of CARMI. First, we build indexes on the OSMC/Face datasets with historical queries that obey a uniform distribution, and query them with 8 different access patterns: seven Zipfian distributions and a uniform distribution to simulate the situation where new sets of keys are queried. To simulate read-write workload shifts, we construct indexes with historical queries under read-heavy and write-heavy workloads, respectively, and test them with write-heavy workloads. Next, we use the uniform dataset to build the index and then perform a write-heavy workload by inserting the keys from the OSMC dataset to simulate the data distribution shifts. As shown in Figure~\ref{fig:drift}, CARMI can still maintain good performance in these situations and is robust to workload and data distribution shifts.}}

\techreport{In this section, we simulate three groups of workload and data distribution shifts to examine the robustness of CARMI.}

\techreport{
\subsubsection{Workload Shift}
First, we examine whether CARMI can cope with read-only workload shifts. We construct indexes from evenly distributed historical queries on OSMC/Face datasets, and query them with 8 different access patterns: seven Zipfian distributions ($f(x) = \frac{1}{x^\alpha} / \sum_{i=1}^N \frac{1}{i^\alpha}$, $\alpha \in \{0.5, 0.75, \ldots, 1.5, 2, 3\}$) and a uniform distribution as a baseline. The average access time of Zipfian indexes decreases with the increase of $\alpha$ and is smaller than that of the baseline, as shown in Figure~\ref{fig:drift}(a). The results show that CARMI is robust and can handle read-only workload shifts well.

Then, we construct indexes with historical queries under read-heavy and write-heavy workloads, respectively, and test them with write-heavy workloads. Although the indexes built on read-heavy workloads require 1.1$\times$ the average time compared to indexes built on write-heavy workloads, as shown in Figure~\ref{fig:drift}(a), CARMI still executes queries quickly and is robust to workload shifts.

\subsubsection{Data Distribution Shift}
Next, we demonstrate the robustness of CARMI to data distribution shifts. We first use the uniform dataset with 67,108,864 key values to build indexes, and then perform a write-heavy workload using 67,108,864 data points from OSMC dataset. Since inserted data points come from a different dataset, we can simulate the situation where the data distribution gradually changes from a uniform distribution to a non-linear distribution. Figure~\ref{fig:drift}(b) shows that although CARMI requires slightly more average time than the index built on OSMC dataset, CARMI can still maintain good performance in this situation. But we still recommend that if data distributions change significantly, reconstruction should be done in time for better performance when sufficient resources are available.

\subsection{Tree Structure}\label{sec:tree}
 To understand the behavior of the index construction algorithm, we have gathered statistics on the optimal indexes constructed by the algorithm, and some of them are shown in Table~\ref{tab:treeStructure}. We follow the convention that the depth of a trivial tree (with only root node) is 1, and count depth on the leaf node level. The average is calculated with respect to keys.
\begin{table}[ht]
	\center
  \small
	\caption{Statistics of Indexes Constructed by CARMI.}
	\label{tab:treeStructure}
	\begin{tabular}{c|cc|cc}
    \toprule
    ~&\multicolumn{2}{c|}{read-only}&\multicolumn{2}{c}{lognormal}\\
    node&uniform &normal&write-heavy&write-partial\\
    \midrule
    root node   & LR & P. LR&P. LR &P. LR \\
    \# of children   &1175653 & 1175653& 1528348&1528348\\
    \midrule
  \# of LR&2& 835&4690&4697\\
  \# of P. LR& 0&4&8&8\\
  \# of Hist& 0&2&2&0\\
  \# of BS&0 &0&0&1\\
    \midrule
    Array& 1175685&1320524&1784066&1767682\\
    \midrule
    Max Depth&3&3&4&4\\
    Avg Depth&2&2.03&2.07&2.07\\
		\bottomrule
	\end{tabular}
\end{table}

The first two columns are statistics under a read-only workload. As we can see, the algorithm constructs an average two-layer structure for uniform dataset under read-only workloads: the first layer is the LR root node and the second layer are the CF array leaf nodes. When performing a read query, we only need to access a single leaf node and a single data block. Due to the distributional variation caused by the sampling process, a small number of data points need an additional inner node.

For other datasets, we can no longer construct a simple two-layer index to handle the dataset as the data points are very concentrated in a specific range, which cannot be handled by a single leaf node. As we can see, the root node setting changes depending on the data distribution and CARMI chooses to use P. LR for all the other three scenarios. We also see that the algorithm generates new inner nodes to partition the data points further, and builds deeper subtrees in more concentrated parts of the data points. Aside from the deep part, the algorithm maintains a shallow free structure in the other parts to guarantee that the average depth is small, achieving an average depth of 2.03-2.07.

As we can see in Table~\ref{tab:treeStructure}, depending on the data distribution, the algorithm will adaptively use the four types of inner nodes. For example, in lognormal dataset under write-partial workload, CARMI constructed $4697$ LR nodes in the evenly distributed part, and $8$ P. LR nodes in the unevenly distributed part. In general, CARMI can effectively handle different data distributions and diverse workloads, and construct a reasonable index structure accordingly. These results also justify the decision of including multiple tree node types in CARMI, which enhances the flexibility of index structures.
} 

\section{Discussion and Future Work}\label{sec:discussion}
\subsection{CARMI in Disk/NVM}
In our current implementation of CARMI, both inner nodes and leaf nodes are stored in memory. We can easily extend CARMI to involve disk operations. If we store data points  on disk, we need to design a new type of leaf node that accesses a disk page instead of an in-memory data block. The corresponding time and space costs of this new type of node should also reflect the latency of disk operations. The current hybrid construction algorithm can be directly used without any change.

Moreover, with the rapid development of non-volatile memory (NVM)~\cite{lankhorst2005low}, the capacity of memory can be vastly expanded while still retaining decently high access speed. The difference in access latency of heterogeneous storage devices can be characterized by modifying the cost model of CARMI, and the construction algorithm will be able to accommodate the new hardware properly.

\subsection{Other Data Types}
In our implementation of CARMI, we only support numerical key types. It is possible to extend CARMI to handle other types as well. Generally speaking, the existing RMI framework cannot readily handle strings since the distribution of strings is hard to fit well by their current models. Our perspective based on data partition can potentially address this difficulty. It is necessary to design new nodes for strings and make certain changes to the storage method. For the hybrid construction algorithm and the overall cost model, we need to make minor changes to them, but ideas and processes of the original algorithm can still be applied.

\techreport{
\subsection{Concurrent Operations}
There have been many methods and implementations on concurrency control of indexes, including B+ Tree~\cite{mohan1989aries, bayer1977concurrency,srinivasan1991performance, graefe2011modern}. The overall structure of CARMI is similar to B+ Tree, and it is also in the form of a tree composed of multiple different nodes. Therefore, the traditional concurrency control method can also be used on CARMI. We can use read/write locks to support concurrency. For lookup operations, we only need to obtain the read locks of the leaf node and its upper layer node when traversing the index. For insert operations, we obtain the write locks on the nodes in the access path. If the leaf node can handle the insert operation without the need for split operation, the lock of the upper nodes can be released.
}

\section{Related Work}\label{sec:related}
In this section, we introduce some related works in the field of database index. \paper{Due to page limitation, some related works, including researches in the field of database that utilize ML techniques~\cite{wang2016database, gani2016survey, kraska2019sagedb, ortiz2018learning, marcus2018towards, krishnan2018learning, kipf2018learned}, are not discussed here and can be found in our technical report~\cite{zhang2021carmi}.}

\textbf{Traditional Indexes: }Many researchers have optimized index structures in the past decades and have proposed many indexes to achieve good performance, such as B-tree, B+ tree~\cite{bayer1970organization,bayer2002organization}, T tree~\cite{lehman1985study}, balanced B-tree~\cite{bayer1972symmetric} and red-black tree~\cite{boyar1994efficient}, etc. Since most indexes are stored in the main memory, CSS-tree~\cite{CSS} restricts node size to the cache line size and eliminates child nodes' pointers to utilize the cache and CSB+ tree~\cite{rao2000making} is then proposed to support updates. ART~\cite{leis2013adaptive} is an adaptive cardinality tree designed to reduce the number of cache misses. For further acceleration, pB+-tree~\cite{chen2001improving} and FAST~\cite{kim2010fast} use prefetching instructions and SIMD instructions, respectively. Masstree~\cite{mao2012cache} effectively handles possible binary keys of any length, including keys with long shared prefixes.

\textbf{Learned Indexes:} Kraska et al.~\cite{kraska2018case} propose a recursive model index (RMI), which uses ML models instead of index structures represented by B-tree. However, RMI does not support inserts. Fiting-tree~\cite{galakatos2018tree} uses linear models to replace the leaf nodes of B-tree to compress the index. PGM-index~\cite{ferragina2020pgm}, a compressed learned index with provable worst-case bounds, extends Fiting-tree and provides an optimal method to find the piecewise linear models. These two indexes support inserts by additional buffers, and the performance needs to be improved. To support writes, ALEX~\cite{ding2019alex} uses model-based inserts and gapped array leaf nodes to make room for future inserts. However, ALEX requires large-range searches to accurately locate data points, resulting in performance degradation on real-world datasets. In this paper, we adopt a data partitioning view and use a cost-based hybrid construction algorithm to select suitable node settings to build an updatable index. CARMI makes full use of various types of nodes to handle different situations, thus maintaining good performance on real-world datasets.

CDFShop~\cite{marcus2020cdfshop} uses Pareto analysis to find Pareto optimal configuration for the static two-layer RMI. They use a parameter search strategy, which is not suitable for dynamic structures due to the exponentially large number of configurations in the search space. On the other hand, CARMI can automatically tune tree structures and model types at runtime through a new dynamic architecture and the cost-based algorithm.

Besides, RadixSpline~\cite{radixspline} focuses more on building indexes in single pass and PLEX~\cite{stoian2021plex} is built on RS and retains only one hyper-parameter for more convenient use. LIPP\cite{wu2021updatable} uses entry types and a conflict degree metric to decide tree node layout. Mitzenmacher et al.~\cite{mitzenmacher2018model} build a learning bloom filter that uses neural network models to predict whether keywords belong to a certain set. In addition, there are several other works that apply the idea of the learned index to multi-dimensional indexes~\cite{ding2020tsunami, nathan2020learning} and spatial query processing~\cite{li2020lisa, wang2019learned, pandey2020case, qi2020effectively} to improve performance.

\techreport{
\textbf{Other Applications of ML in Database:} In other parts of the database, many novel ideas that utilize ML techniques have also been proposed. Park et al.~\cite{park2017database} applied ML techniques to approximate query processing. They found that the answer to each query revealed some degree of knowledge about the answer to another query. Because their answers come from the same underlying distribution that produced the entire dataset. Therefore, it is possible to use ML techniques to improve this knowledge and answer questions in a more organized manner instead of reading a lot of raw data. In addition, several research teams~\cite{ortiz2018learning, marcus2018towards, krishnan2018learning, kipf2018learned, sun2021learned, wu2021unified} proposed the use of reinforcement learning, CNN, DNN and other AI related models to predict cardinality and optimize queries.

Cumin et al.~\cite{cumin2017data} proposed a rewriting technique of SQL query based on ML. This technology can help data scientists formulate SQL queries and browse big data quickly and intuitively. At the same time, the user input can be minimized without manual tuple description or marking.

Kraska et al. proposed a blueprint of a database that uses ML techniques in each part~\cite{kraska2019sagedb, kraska2021ml} to improve performance and discussed several ideas for designing new benchmarks~\cite{bindschaedler2021towards} for learned systems. ADCP~\cite{ma2020active}, an active data collection platform, and HAL~\cite{ma2020active}, a novel active learning technique to gather data in the database, are applied to handle the problem of performance degradation of ML models in use.
}

\section{Conclusion}\label{sec:conclusion}
This paper conducts in-depth research on the basic framework of learned indexes (RMI), and argues that the inflexibility of data partitioning is an important reason for the performance degradation of RMIs in practical scenarios. To address this issue, we propose to view RMI construction from a data partitioning view and propose a general cache-aware RMI framework, called CARMI. Specifically, we use the entropy metric to quantify the data partitioning effectiveness and propose a new cost model to characterize the performance of individual tree nodes, which helps to design more robust indexes. Furthermore, CARMI is equipped with a new memory layout that is more cache-friendly and uses a hybrid algorithm to automatically construct index structures for different datasets and workloads without manual tuning. Experimental results show that CARMI has an outstanding performance under a variety of datasets and workloads, achieving an average of $2.2\times$/$1.9\times$ speedup compared to B+ Tree/ALEX, respectively, while using only $0.77\times$ the memory space on average. This paper demonstrates that data partitioning is important for learned indexes during construction and that the new cost model can well characterize the performance of a single node, which is beneficial to improving the performance and flexibility of indexes. Finally, the CARMI framework is highly extensible and robust, and can be applied to a wider range of scenarios if we design different types of nodes for it.

\bibliographystyle{ACM-Reference-Format}
\bibliography{sample-base}
\techreport{
  \appendix \label{sec:appendix}
\section{Details of Inner Nodes}\label{sec:innerdetails}
\subsection{Linear Regression Node}
Linear regression node (LR node) is the type of inner node that uses linear regression to determine the next branch. As mentioned in Section~\ref{sec:cache-aware}, the size of tree nodes must be 64 bytes. We only use a simple linear model in this node to achieve the goal of predicting the position as quickly as possible, and the remaining bytes are placeholders. For a given input key, the next branch can be directly obtained by a few simple calculations and boundary processing. Figure~\ref{fig:lrStructPro} shows the memory layout and detailed computation steps of LR nodes for reference.
\begin{figure}[h]
  \vspace{-0.2cm}
  \centering
  \includegraphics[width=0.9\linewidth]{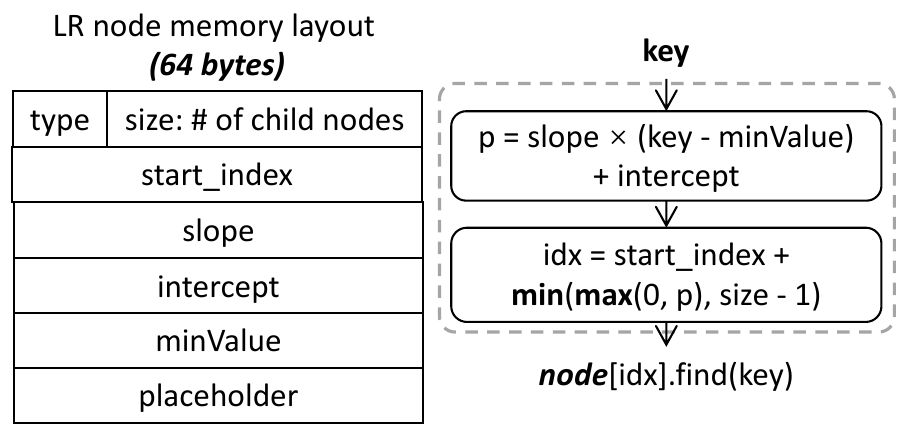}
  \caption{Details of LR Node}
  \Description{Details of LR Node}
  \label{fig:lrStructPro}
\end{figure}

\subsection{Piecewise Linear Regression Node} \label{sec:nn introduction}
Piecewise linear regression node (P. LR node) uses a piecewise linear regression model to determine the next branch. More specifically, the piecewise linear model contains {$kIndexNum + 1$} segments, and we store the coordinates of the segment end points in the P. LR node. Due to the fixed number of segments, the training method of the PLR node, which is related to the segment point, is slightly different from the least-squares method of the LR node. $kIndexNum$ is calculated from the size of the key type to ensure that the total size does not exceed 64 bytes, and placeholders fill the part less than 64 bytes. For a 32-bit key, its value is 6, which means that there are a total of 8 linear models. When we traverse through a P. LR node, we first find the segment to which the input key belongs, then use the segment as a linear model to determine the next branch. The memory layout and internal mechanism of a P. LR node are demonstrated in Figure~\ref{fig:plrStructPro}.
\begin{figure}[h]
  \vspace{-0.2cm}
  \centering
  \includegraphics[width=0.9\linewidth]{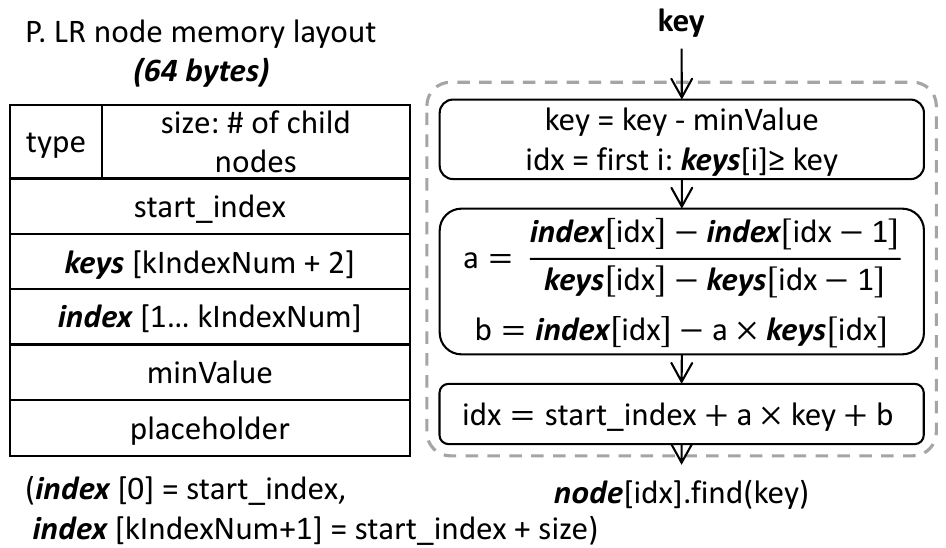}
  \caption{Details of P. LR Node}
  \Description{Details of P. LR Node}
  \label{fig:plrStructPro}
\end{figure}

\subsection{Binary Search Node}
Binary search node (BS node) uses a binary search procedure to determine the corresponding branch for the input key value. This is very similar to traversing through B+ Tree, the only difference is that the reference table in a BS node is determined during the index construction procedure, and remains fixed afterward.
\begin{figure}[htbp]
  \centering
  \begin{minipage}[t]{0.38\linewidth}
  \centering
  \includegraphics[width=3.5cm]{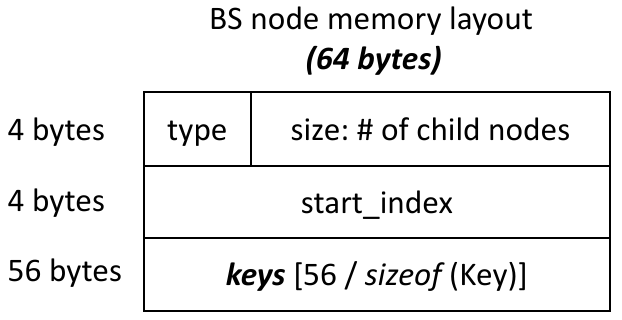}
    \Description{Structure of LR node}
  \end{minipage}
  \begin{minipage}[t]{0.6\linewidth}
  \centering
  \includegraphics[width=5cm]{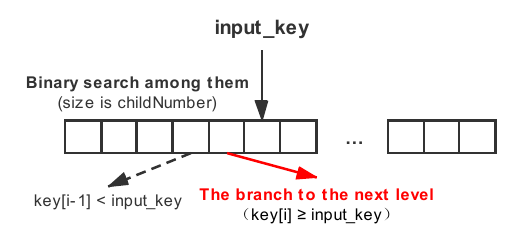}
    \Description{Data access process of BS node}
  \end{minipage}
  \caption{Details of BS Node}
  \label{fig:bsStructPro}
\end{figure}

Each BS node stores {$\frac{56}{sizeof(key)}$} key values internally, and divides the key range into $\frac{56}{sizeof(key)} + 1$ intervals. To determine which branch to go through, a binary search is performed among the stored key values to locate the corresponding key value interval covering the input key. Figure \ref{fig:bsStructPro} shows the details of the BS nodes.

\subsection{Histogram Node} \label{sec:histogram introduction}
Histogram nodes find the corresponding branch directly through a simple one-step calculation and table look-up operation.
\begin{figure}[h]
  \vspace{-0.2cm}
  \centering
  \includegraphics[width=0.9\linewidth]{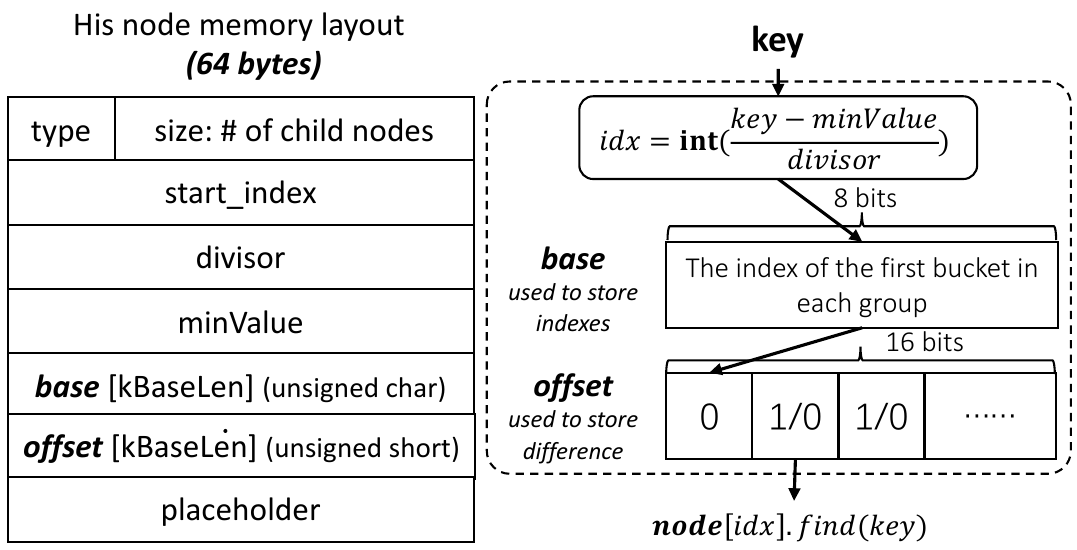}
  \caption{Details of Hist Node}
  \Description{Details of Hist Node}
  \label{fig:hisStructPro}
\end{figure}

The look-up table in Histogram node is stored very compactly using two tables. Each byte in \textit{Base} represents the index in child nodes of the first bit of the corresponding 16 bits in \textit{Offset}. Each bit of \textit{Offset} represents the difference (0 or 1) between the index of the current bucket and the previous bucket, as shown in Figure~\ref{fig:hisStructPro}. To determine the next branch,  we only need to visit \textit{Base} to get the base index, and then count the number of bits in the \textit{Offset} table, and finally add them together to get the index of the next branch.

\section{Details of Leaf Nodes} \label{sec:details of leaf nodes}

\subsection{CF Array Leaf Node}
In cache-friendly array leaf nodes (CF node), the data points are compactly stored in data blocks in a sequential manner. When searching for data points in the CF array leaf nodes, we first search sequentially in the keywords of each block, and then search in the block. When we need to insert a given data point, we must first find the correct position, and then move all the data points by one cell to make room for the new data point. The memory layout and internal mechanism of a CF leaf node are demonstrated in Figure~\ref{fig:cf}.
\begin{figure}[h]
  \vspace{-0.2cm}
  \centering
  \includegraphics[width=\linewidth]{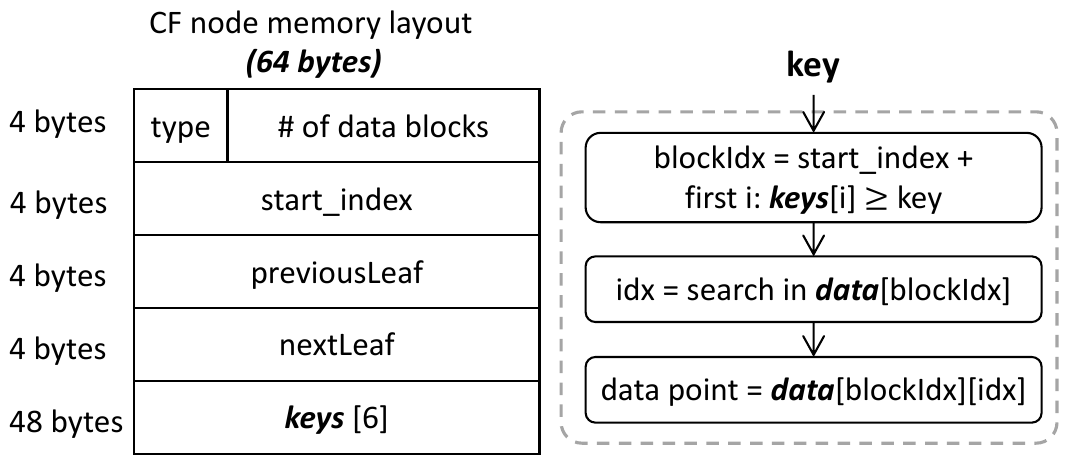}
  \caption{Details of CF Nodes}
  \Description{Details of CF Nodes}
  \label{fig:cf}
\end{figure}

\subsubsection{Insert Operation}
In Section~\ref{sec:insert}, we have introduced the procedure of insert operations with two general mechanisms to make room for the data points to be inserted. In the implementation of CARMI, we actually use three mechanisms in the CF leaf node: rebalance, expand, and split. Figure~\ref{fig:insert} shows the mechanisms of them. Here we explain the rebalance mechanism that has not been introduced.
\begin{figure}[h]
  \vspace{-0.2cm}
  \centering
  \includegraphics[width=\linewidth]{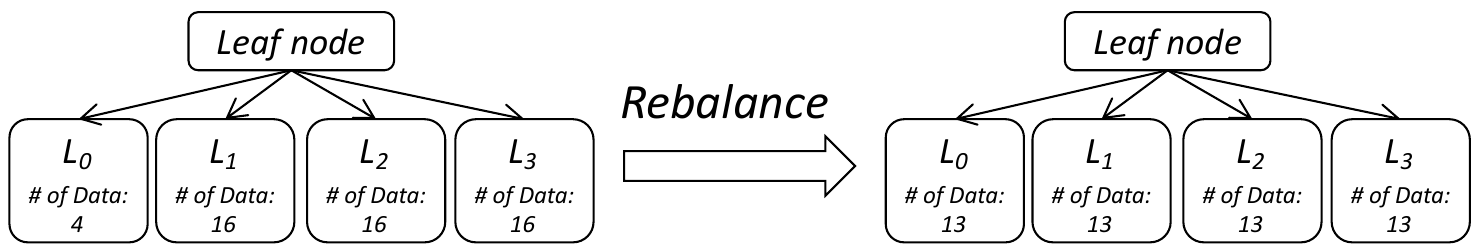}
  \caption{Rebalance Mechanism}
  \Description{Rebalance Mechanism}
  \label{fig:insert}
\end{figure}

When the data block to be inserted is full, but the next data block is not full, we will insert the data point into the next block. In order to keep all data points managed by this leaf node in order, the inserted data point will be compared with the largest one in the current block. The larger one is inserted into the next block, and the other is stored in the current block. If the next data block is also saturated, the rebalance mechanism is triggered. The leaf node collects all data points in its data blocks and reallocates them evenly to all these blocks. This mechanism can make the data distribution in data blocks more even, reducing a certain amount of time for subsequent insert.

\subsection{External Array Leaf Node} \label{sec:extleaf}
The external array leaf node uses a simple linear model to predict the location, and the pointer no longer points to the data array but to the external location. The external array leaf node is used to create an index over a sorted array of data points (e.g., primary indexes in databases), and we use this type of node to create an index for the YCSB dataset. Its access flow is shown in Figure~\ref{fig:tmpLeaf}.

Inevitably, the accuracy of the model prediction cannot reach 100\%, so we need a subsequent binary search step to find the exact position of the data record. For this purpose, we employ an additional parameter: the error $\varepsilon$, whose value is determined during the training process. For a given key value $x$, let us denote the actual location of the data point as $loc(x)$, and the predicted location of the data point as $preIdx(x)$. Then, when looking for the data point matching $x$, depending on the relationship between these two terms, there are essentially two cases: either $loc(x) \in [preIdx(x) - \frac{\varepsilon}{2}, preIdx(x) + \frac{\varepsilon}{2}]$, or $loc(x)$ is outside of this range. The time cost of data access is also different for these two cases. In the former case, we need to perform a binary search over the range of $[preIdx(x) - \frac{\varepsilon}{2}, preIdx(x) + \frac{\varepsilon}{2}]$, while in the latter case the range of the binary search is the entire array. 
\begin{figure}[h]
  \vspace{-0.2cm}
  \centering
  \includegraphics[width=\linewidth]{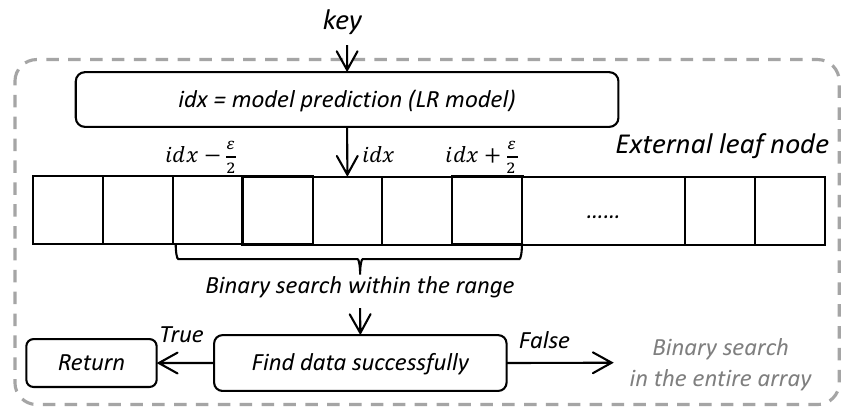}
  \caption{Details of External Array Leaf Nodes}
  \Description{Details of External Array Leaf Nodes}
  \label{fig:tmpLeaf}
\end{figure}

\subsubsection{The Optimal Value of Error Parameters}\label{sec:error parameter}

Essentially, we want to choose the error parameter $\varepsilon$ that minimizes the average time cost of data access. The method for finding its optimal value is given below. Let us denote for each data point its true position as $y_i$, the predicted value as $p_i$, and their difference $d_i = y_i - p_i$. With these notations, we can compute the average time cost of data access as in Equation~(\ref{eq:error}).
\begin{equation}\label{eq:error}
Avg = \frac{1}{n} [|\{i:|d_i| \leq \frac{\varepsilon}{2}\}|\cdot  \lfloor \log_2 \varepsilon \rfloor + |\{i:|d_i| > \frac{\varepsilon}{2}\}| \cdot \lfloor \log_2 n \rfloor]
\end{equation}

Then, we only need to substitute different $\varepsilon$ values into Equation~(\ref{eq:error}) to find the minimum value of $Avg$, where the range of $\varepsilon$ is from 0 to $\max\left| p_i - y_i \right|$.
$$ \varepsilon = \arg\min_\varepsilon |\{i:|d_i| \leq \frac{\varepsilon}{2}\}|\cdot  \lfloor \log_2 \varepsilon \rfloor \nonumber \\ + |\{i:|d_i| > \frac{\varepsilon}{2}\}| \cdot \lfloor \log_2 n \rfloor $$

\begin{table*}[h]
	\center
	\caption{The Settings of Parameters} \label{tab:para}
	\begin{tabular}{ccc}
    \toprule
    Params &Explanation&Default Value \\
    \midrule
    $\lambda$&several different values were used in experiments & 0.03 \\
    \midrule
    $kBlockSize$& the fixed size of a data block & 256\\
    \midrule
    \multirow{2}*{$kLeafMaxCapacity$} & (1) the maximum capacity of CF nodes& 96\\
    ~ & (2) the maximum capacity of external nodes & 512\\
    \midrule
    $kDPThreshold$ & used to switch between DP and greedy algorithms& 512\\
    \midrule
    $kDividedNodeNum$& the number of new leaf nodes in split mechanism & 16\\
    \midrule
     $kLeafThreshold$ & the size threshold in DP transition & 90\\
		\bottomrule
	\end{tabular}
\end{table*}

\section{Details of Root Nodes}
\subsection{Linear Regression Root Node}\label{sec:lrroot}
The linear regression root node is a very simple root node consisting of a linear regression model. In addition to the basic members, only two linear model parameters need to be stored. We only need one model prediction and the boundary condition processing to get the index of the next node. The details of LR root node is shown in Figure~\ref{fig:LRRoot}.
\begin{figure}[h]
  \vspace{-0.2cm}
  \centering
  \includegraphics[width=\linewidth]{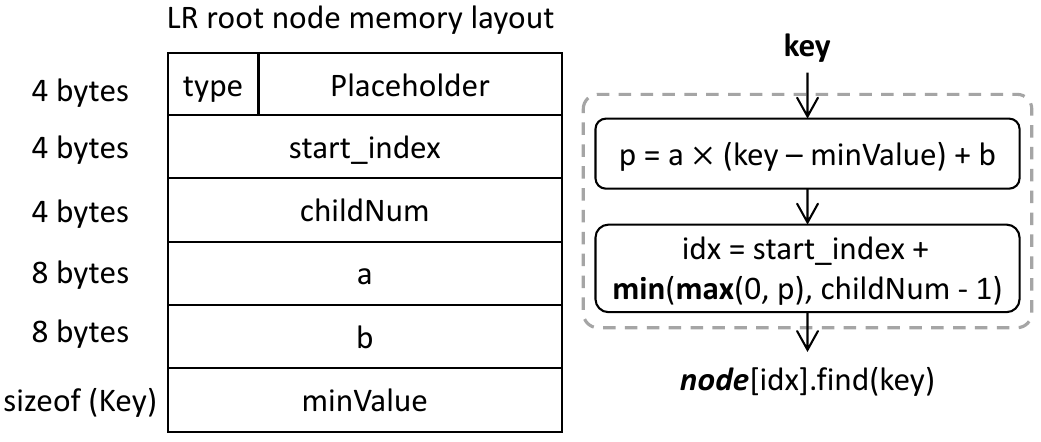}
  \caption{Details of LR Root Nodes}
  \Description{Details of LR Root Nodes}
  \label{fig:LRRoot}
\end{figure}

\subsection{Piecewise Linear Regression Root Node}\label{sec:plrroot}
The P. LR root node uses a piecewise linear regression model with up to five segments to allocate the dataset to the child nodes. It differs from the P. LR inner nodes only in the number of segments. We use a dynamic programming algorithm similar to the prefetch model to find the optimal segmentation points and the number of segments, thereby maximizing the entropy of the node. Since the root node is in the cache, we do not limit its size here. For the P. LR root node, we find the first breakpoint greater than or equal to the key value, and then use the corresponding model parameters for calculation and boundary processing, as shown in Figure~\ref{fig:PLRRoot}.
\begin{figure}[h]
  \vspace{-0.2cm}
  \centering
  \includegraphics[width=\linewidth]{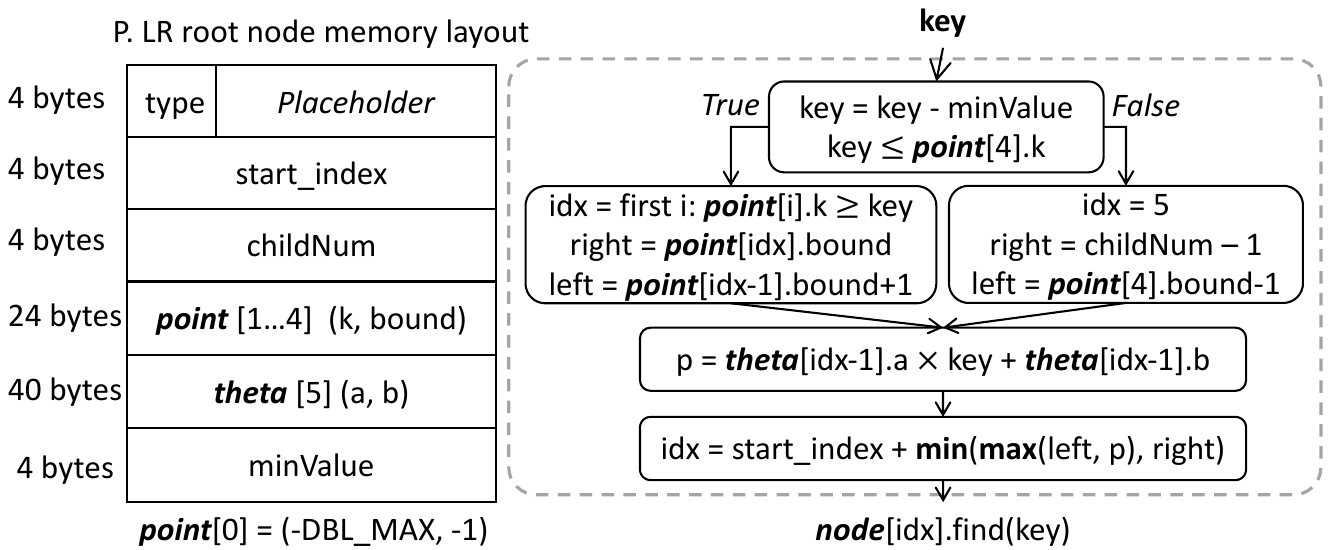}
  \caption{Details of P. LR Root Nodes}
  \Description{Details of P. LR Root Nodes}
  \label{fig:PLRRoot}
\end{figure}

\section{Parameter Settings} \label{sec:para setting}
This section describes the default parameter settings in our experiments. The default value of the parameter $\lambda$ used to tradeoff between the time cost and space cost is 0.03, but we adjusted this parameter in the experiments according to the datasets and the needs of the experiments. The size of each data block is represented as $kBlockSize$, and the default value is 256 bytes, and the maximum capacity of external leaf nodes is 512. The parameter $kDPThreshold$ is used to switch between the dp and greedy algorithms and its value is $512$. When a leaf node needs to perform a split operation, we need to replace it with an inner node and $kDividedNodeNum$ leaf nodes with a default value of 16. The parameter $kLeafThreshold$ is $90$, which means that when the number of data points is less than $90$, the algorithm will directly construct a leaf node instead of choosing a better one from leaf nodes and inner nodes, thus saving a certain amount of space cost and construction time. Note that this parameter must be smaller than $kLeafMaxCapacity$. Table~\ref{tab:para} lists these parameters with default values for a quick view.

\section{Proof of Theorem~\ref{theo:entropy}}
\begin{proof}
W.l.o.g, let $M_1, \ldots, M_{I}$ be the set of inner nodes, and $M_{I+1}, \ldots, M_{K}$ be the set of leaf nodes. Let $n_i$ be the number of data points in node $M_i$, and $C_i$ be the set of child nodes of $M_i$. Then, we can substitute the definition of $P(M_i)$ and $H(M_i)$ into Theorem~\ref{theo:entropy}:
  \begin{equation*}\label{proof:entropy}
    \begin{split}
      \textstyle
       \sum_{i=1}^K P(M_i)H(M_i)& = -\sum_{i=1}^{I} \frac{n_i}{n} \sum_{j \in C_i} \frac{n_j}{n_i} \log_2 \frac{n_j}{n_i} + \sum_{i={I+1}}^{K} \frac{n_i}{n} \log_2 {n_i}\\
      & = -\sum_{i=1}^I \sum_{j \in C_i} \frac{n_j}{n} \log_2 \frac{n_j}{n_i} + \sum_{i = I+1}^K \frac{n_i}{n}\log_2 n_i\\
    \end{split}
  \end{equation*}
  
For each leaf node $M_l$, let the traverse path from the root to $M_l$ be $Root \rightarrow M_{i_1} \rightarrow M_{i_2}  \rightarrow \ldots \rightarrow M_{i_k} \rightarrow M_l$. We can define $P_i$ as the following set: $P_i = \left\{ (Root, M_{i_1}), (M_{i_1}, M_{i_2}), \ldots (M_{i_k}, M_l)	\right\}$. Then, the double summation can be rearranged according to all the paths from the root node to the leaf nodes:
\begin{equation*}
  \begin{split}
    \textstyle
    -\sum_{i=1}^I \sum_{j \in C_i} \frac{n_j}{n} \log_2 \frac{n_j}{n_i} &= \sum_{i=1}^I \sum_{j \in C_i} \sum_{l:(M_i, M_j) \in P_l} \frac{n_l}{n} \log_2 \frac{n_i}{n_j}\\
  \end{split}
\end{equation*}
where we split $\frac{n_j}{n}$ according to whether the $l$-th leaf node is visited to obtain $\sum_{l:(M_i, M_j) \in P_l} \frac{n_l}{n}$.

Then we sort and rearrange them according to each path to the leaf node:
\begin{equation*}
  \begin{split}
    \textstyle
    -\sum_{i=1}^I \sum_{j \in C_i} \frac{n_j}{n} \log_2 \frac{n_j}{n_i} & =\sum_{l={I+1}}^{K}\frac{n_l}{n} \sum_{(M_i, M_j) \in P_l} \log_2 \frac{n_i}{n_j}\\
  \end{split}
\end{equation*}

Finally, we merge the above equation with the leaf node part
\begin{equation*}
  \begin{split}
    \textstyle
    \sum_{i=1}^K P(M_i)H(M_i)
    & =\sum_{l={I+1}}^{K}\frac{n_l}{n} ( \sum_{(M_i, M_j) \in P_l}   \log_2 \frac{n_i}{n_j} + \log_2 {n_l})\\
  \end{split}
\end{equation*}
and simplify the equation in the path order:
\begin{equation*}
  \begin{split}
    \textstyle
    \sum_{i=1}^K P(M_i)H(M_i)
    & =\sum_{l={I+1}}^{K}\frac{n_l}{n} (\log_2 \frac{n}{n_l} + \log_2 {n_l})\\
    & =\sum_{l={I+1}}^{K}  \frac{n_l}{n} \log_2 n\\& = \log_2 n\\
  \end{split}
\end{equation*}
where $\sum_{l={I+1}}^{K} n_l$ is equal to $n$.
\end{proof}

}
\end{document}